\documentstyle[11pt,aaspp4,epsfig]{article}

\def\gvel{$\gamma^{2}$\,Vel}
\def\zpup{$\zeta$\,Pup}
\def\dori{$\delta$\,Ori~A}
\def\pap1{Paper~I}
\def\fuse{{\it FUSE\/}}
\def\iue{{\it IUE\/}}
\def\hst{{\it HST\/}}
\def\oao{{\it Copernicus}}
\def\orf{{\it ORFEUS-SPAS II}}
\def\lya{Ly$\alpha$}
\def\lyb{Ly$\beta$}
\def\lyg{Ly$\gamma$}
\def\lyd{Ly$\delta$}
\def\lye{Ly$\epsilon$}

\def\EE#1{\times 10^{\small#1}}
\def\cmsq{\rm ~cm^{\small -2}}
\def\cm3{\rm ~cm^{\small -3}}
\def\kms{\rm ~km~s^{\small -1}}

\def\ergcmsA{\rm\,erg~cm^{\small -2}~s^{\small -1}~\AA^{\small -1}}

\def\wl{$\lambda$}
\def\wll{$\lambda\lambda$}
\def\h2{H$_2$}
\def\hi{{\ion{H}{1}\,}}
\def\hii{{\ion{H}{2}\,}}
\def\di{{\ion{D}{1}\,}}
\def\nhi{$N$({\ion{H}{1})\,}}
\def\ndi{$N$({\ion{D}{1})\,}}
\def\nni{$N$({\ion{N}{1})\,}}
\def\noi{$N$({\ion{O}{1})\,}}

\def\ni{{\ion{N}{1}}}

\def\oi{{\ion{O}{1}}}

\def\feii{{\ion{Fe}{2} }}
\def\chisq{$\chi^{2}$}

\def\nav{$N_a(v)$}

\slugcomment{}
\lefthead{SONNEBORN ET AL.}
\righthead{DEUTERIUM TOWARD $\gamma ^{2}$ VEL AND $\zeta$ PUP}

\begin{document}
\title{Spatial Variability in the Ratio of Interstellar Atomic 
Deuterium to Hydrogen. II. Observations toward $\gamma ^{2}$ Velorum
and $\zeta$ Puppis  by the Interstellar Medium Absorption Profile 
Spectrograph\footnote{This paper is dedicated in memory of Judith L. Tokel, wife of the first author, who passed away on 2000 June 10. Her enthusiastic support and encouragement were essential to its successful completion.}}

\author{George Sonneborn,\altaffilmark{2,3}
Todd M. Tripp,\altaffilmark{4}
Roger Ferlet,\altaffilmark{3,5}
Edward B. Jenkins,\altaffilmark{4}
U. J. Sofia,\altaffilmark{6}
Alfred Vidal-Madjar,\altaffilmark{3,5} and
Prezemys{\l}aw R. Wo\'zniak\altaffilmark{4}}
\affil{}

\altaffiltext{2}{Laboratory for Astronomy and Solar Physics, Code 681,
NASA Goddard Space Flight Center, Greenbelt, MD 20771; 
~george.sonneborn@gsfc.nasa.gov}
\altaffiltext{3}{Guest Investigator with the IMAPS instrument on the 
{\it ORFEUS-SPAS II} mission}
\altaffiltext{4}{Princeton University Observatory, Princeton, NJ 08544;
~tripp@astro.princeton.edu, ebj@astro.princeton.edu, wozniak@astro.princeton.edu}
\altaffiltext{5}{Institut d'Astrophysique de Paris, 98bis Blvd. 
Arago, 75014 Paris, France; ~ferlet@iap.fr, alfred@iap.fr}
\altaffiltext{6}{Dept. of Astronomy, Whitman College, 345 Boyer Ave., 
Walla Walla, WA 99362; ~sofiauj@whitman.edu}

\begin{abstract}

 High resolution far 
ultraviolet spectra of the early-type stars \gvel\ and \zpup\  were 
obtained to measure the interstellar deuterium abundances in these directions. The observations were made with the Interstellar Medium Absorption Profile Spectrograph (IMAPS) 
during the {\it ORFEUS-SPAS II} mission in 1996.  IMAPS spectra cover 
the wavelength range ~930--1150\AA\ with $\lambda/\Delta\lambda\sim 80,000$.  
The interstellar \di\ features are resolved and cleanly separated from 
interstellar \hi\ in the \lyd\ and \lye\ profiles of both sight lines, and 
also in the \lyg\ profile of \zpup.  The \di\ profiles were modeled using a 
velocity template derived from several \ni\ lines in the IMAPS spectra 
recorded at higher signal-to-noise.   To find the best \di\ column density,
we minimized \chisq\ for model \di\ 
profiles that included not only the \ndi\ as a free parameter, 
but also the effects 
of several potential sources of systematic error which were allowed to vary 
as free parameters.  \hi\ column densities were measured by analyzing  \lya\ 
absorption profiles in a large number of \iue\ high 
dispersion spectra for each of these stars and applying this same 
\chisq-minimization technique.  
Ultimately we found that D/H $= 2.18^{+0.36}_{-0.31} \EE{-5}$ for \gvel\ and 
$1.42^{+0.25}_{-0.23} \EE{-5}$ for \zpup, values that contrast  markedly with 
D/H derived in \pap1\ for \dori\ (the stated errors are 90\% confidence limits).
Evidently, the atomic D/H ratio in the ISM, averaged over path lengths of 
250 to 500 pc, exhibits significant spatial variability. Furthermore, the 
observed spatial variations in D/H do not appear to be anticorrelated with N/H, 
one measure of heavy element abundances. We briefly discuss some 
hypotheses to explain the D/H spatial variability. Within the framework of standard Big Bang Nucleosynthesis, the large value of D/H found toward \gvel\ is equivalent to a cosmic baryon density of $\Omega_B h^2 = 0.023\pm0.002$, which we regard as an upper limit since there is no correction for the destruction of deuterium in stars.

\end{abstract}

\keywords{ISM: Abundances --- Cosmology: Observations --- ISM: Evolution --- 
Ultraviolet: ISM --- Stars: Individual (\gvel, \zpup)}

\section{INTRODUCTION}

The abundance ratio of atomic deuterium to hydrogen (D/H) in interstellar gas is widely regarded as an important tracer of galactic chemical evolution 
(Audouze \& Tinsley 1974; Boesgaard \& Steigman 1985; Tosi et al. 1998) 
and a key discriminant 
of the cosmic baryon-to-photon ratio $(\eta)$ in Big Bang Nucleosynthesis
(BBN, Walker et al. 1991). 
A standard interpretation is that D should not be produced in significant 
quantities in astrophysical sites other than the Big Bang (Epstein, Latimer, \& 
Schramm 1976). The generally accepted viewpoint is that while D is produced 
primordially, its destruction takes place when some of the gas is cycled through 
stars.  The uncertainties surrounding this process represent a
stumbling block in arriving at the primordial value.  For this reason, a
measurement of D/H is often regarded as a lower limit to the primordial
ratio, and this may be translated into an upper limit to $\eta$ (for
BBN, larger D/H implies lower $\eta$).

Rogerson \& York (1973) made the first measurement of 
the atomic D/H abundance ratio in the interstellar medium (ISM) on the line of 
sight toward $\beta$ Cen.   Measurements with the \oao\ satellite toward an 
additional 14 stars ($100< d < 500$ pc) found that D/H values  cluster around 
$1.5\times 10^{-5}$, but with a dispersion that, in
some cases, seemed to exceed the stated uncertainties (see review by
Vidal-Madjar \& Gry 1984).  While a simple interpretation of the \oao\
data suggested that the D/H measurements revealed spatial variations,
the reality of these differences has been difficult to substantiate due to 
concerns stemming from the somewhat inadequate resolution of \oao\ 
 ($15 \kms $ FWHM) for this purpose.

Using the Goddard High Resolution Spectrograph (GHRS) and Space Telescope 
Imaging Spectrograph (STIS) on the {\it Hubble Space Telescope} (\hst), several 
measurements of D/H in the local interstellar medium (LISM, $d<100$ pc) have 
been made with observations of \lya\ (see Lemoine et al. 1999 for a review). 
Linsky et al. (1993, 1995) found D/H $=1.60^{+0.14}_{-0.19} \EE{-5}$ toward 
Capella. Vidal-Madjar et al. (1998)  reported evidence  for a factor of 
$\sim2$ difference in D/H between two components on the line of sight toward 
the white dwarf G191-B2B ($d=75$ pc).  Sahu et al. (1999) re-evaluated the GHRS 
data as well as new STIS echelle spectra of G191-B2B and concluded that, as a 
result of improved instrumental characterization (echelle scattered light 
correction near \lya), D/H values in both components appeared to be consistent with the 
usual LISM value. Questions remain, however. For example, Howk \& Sembach (2000) 
found a different STIS echelle scattered light correction, one that agrees with 
the  GHRS \lya\ profile (Vidal-Madjar 2000). The resolution of the 
conflicting conclusions about the G191-B2B sightline may be provided by 
observations of higher Lyman series lines (\lyb\ and \lyg) by the {\it Far 
Ultraviolet Spectroscopic Explorer} (\fuse, Moos et al. 2000).

The Interstellar Medium Absorption Profile Spectrograph (IMAPS) provides a new 
window to study deuterium in the Galaxy by virtue of its wavelength coverage and 
very high spectral resolution ($\lambda/\Delta\lambda\sim 80,000$) which 
alleviate many of the previous obstacles to accurate D/H measurements. Prior to 
the IMAPS deuterium studies, D/H measurements in our galaxy after the \oao\ 
mission have been confined to the local ISM, since the saturated core of \hi 
\lya\ envelopes the \di\ \lya\ feature for \nhi$\ga 1\EE{19} \cmsq $ and higher 
Lyman lines lie beyond the reach of \hst. The goal of the IMAPS deuterium 
program is to obtain high-quality measurements of D/H on sightlines toward 
bright OB stars beyond the LISM. In the first paper of this series, Jenkins et 
al. (1999a, hereafter \pap1)  measured the D/H abundance 
ratio toward \dori\  using IMAPS spectra. They found D/H $=0.74^{+0.19}_{-0.13}\EE{-5}$, a value that 
is  much lower than that found in the  LISM.  \pap1\
 also showed that the low abundance of D toward \dori\ is {\it not} 
accompanied by an overabundance of N and O relative to H, as would be expected 
if the gas had been more thoroughly cycled through stellar interiors. We note 
that two other deuterium measurements using the T\"ubingen echelle spectrograph 
(resolution $\lambda/\Delta\lambda\sim 10,000$) on the \orf\ mission have been 
recently reported: G\"olz et al. (1998) found D/H $=0.8^{+0.7}_{-0.4}\EE{-5}$ toward 
BD\,+28$^{\circ}$\,4211, an O-type subdwarf $\sim$100 pc away, and Bluhm 
et al. (1999) measured D/H $=1.2^{+0.5}_{-0.4}\EE{-5}$ toward 
BD\,+39$^{\circ}$\,3226, a sdO star located $\sim270$ pc from the Sun. 

In this paper we analyze the lines of sight toward two well-known hot stars, 
\gvel orum (HD 68273) and \zpup pis (HD 66811). They are among the brightest 
stars in the sky at 1000 \AA, well studied by \oao, have \hi\ column densities 
$\la 10^{20} \cmsq$, and hence are prime D/H targets for IMAPS.  These stars 
were observed under an \orf\ Guest Investigator program.

\placetable{stardata}
\begin{deluxetable}{lcccccccc}
\tablewidth{0pc}
\tablecaption{Stellar data\label{stardata}}
\tablehead{ Name & HD & $l^{II}$ & $b^{II}$ & Spectral & $V$ & 
  v$\sin\,i$ & $\Delta v_{LSR}$\tablenotemark{a} & $d$\tablenotemark{b} \nl
~ & ~ & ~ & ~ & Type & (mag) & ($\kms$) & ($\kms$) & (pc)  }
\startdata
\zpup\ & 66811 & 253\fdg98 & -4\fdg71 & O5~Iaf & 2.25 & 211 & -18.04 & 
$429^{+120}_{-77}$ \nl
\gvel\ & 68273 & 262\fdg81 & -7\fdg70 & WC8+O8 & 1.78 & \ldots & -17.48 & 
$258^{+41}_{-31}$ \nl
\enddata
\tablenotetext{a}{~$v_{LSR} = v_{helio} + \Delta v_{LSR}$}
\tablenotetext{b}{Schaerer et al. (1997)}
\end{deluxetable}
\clearpage
The sightlines to \gvel\ and \zpup\ have been studied extensively (\gvel: Morton 
\& Bhavsar  1979; Sahu 1992; Fitzpatrick \& Spitzer 1994; \zpup: Morton \& 
Dinerstein 1976; Morton 1978), including D/H measurements with observations from 
the \oao\ satellite. York \& Rogerson (1976) found D/H$=2.0^{+1.1}_{-0.7}\EE{-
5}$ toward \gvel.  Vidal-Madjar et al. (1977) studied the sightline toward 
\zpup\ and found that solutions consistent with the data permit a wide range of D/H values  ($1.7 \EE{-5} \la$ D/H $\la 2.5 \EE{-4}$).
\gvel\ and \zpup\ are the most luminous stars in the Vela-Puppis region. Their 
role in powering the Gum Nebula and their relationship to the Vela OB2 
association, Vela supernova remnants, and various structures in the ISM in Vela-
Puppis have been analyzed by Sahu (1992). Table \ref{stardata} summarizes some 
basic properties of the two stars. The distances listed were derived from {\it 
HIPPARCOS} data  by Schaerer, Schmutz \& Grenon (1997), placing \gvel\ 
significantly closer than earlier estimates. 

\gvel\ is the closest (and brightest) Wolf-Rayet star and 
consequently it has been studied intensively for more than a century 
(van der Hucht et al. 1981). Nevertheless, due to 
the complexity of the stellar system, its fundamental stellar 
parameters have been continuously debated and revised (as recently as 
1997 when several papers were published, e.g., Schaerer et al. 1997; Schmutz et 
al. 1997). \gvel\ is a double-lined spectroscopic binary star composed of a WR 
star (WC8) and a late O star (O8 III) of comparable brightness. The complex 
properties of this system are evident in the significant changes in the UV 
spectrum, and this behavior will be addressed in \S\ref{gvelN(HI)} when we 
discuss our determination of the interstellar hydrogen column density through 
measurements of the \lya\ absorption feature.

\zpup\ is an extremely luminous, massive star (O4 Iaf) and the brightest O-type 
star in the sky. It is a prototype for O-star spectral properties and 
variability (Morton \& Underhill 1977; Massa et al. 1995) and theoretical 
understanding of radiation-driven winds in massive stars (e.g. Puldrach et al. 
1994).  \zpup\ is believed to be a single star.

The high velocity resolution provided by IMAPS opens the
way for more accurate line profile measurements in the far ultraviolet spectra 
of bright stars.  Our observations of \gvel\ and \zpup\ described in \S\ref{obs} build on our earlier results for \dori\ described in \pap1.  
We follow with discussions of
how we derived \di\ column densities (\S\ref{N(DI)}), \hi\ column densities
(\S\ref{N(HI)}), and  values for D/H and N/H
(\S\ref{D/H}). 
The paper concludes with a discussion in \S\ref{discuss} of the
possible implications of the differences in D/H for the three stars
covered in this paper and \pap1.

\section{OBSERVATIONS}\label{obs}

The spectra analyzed in this paper were obtained with IMAPS during the \orf\ 
mission (STS-80) in late 1996 (see Hurwitz et al. 1998).
The design of IMAPS and its performance on earlier flights have been described in detail by 
Jenkins et al. (1988, 1996). In \pap1\ we described the  in-flight performance  of IMAPS  during the 1996 mission.
Here we only summarize the instrument's principal characteristics. 

IMAPS 
is an objective-grating echelle spectrograph designed to record the far-UV 
spectrum of bright stars at high spectral resolution.  The optical design 
consists of a wire grid collimator to reject off-axis light from stars other 
than the target, an echelle grating with a 63\arcdeg\ blaze angle, and a 
parabolic cross disperser. The spectral format is imaged on a solid KBr 
photocathode, whose electrons are magnetically focussed on a windowless, 
back-illuminated  CCD with a 320x256 pixel format.  
The nominal wavelength range of 
930--1150 \AA\ is obtained in four selectable tilts that 
span the free spectral range of the echelle grating. The gratings were coated with LiF over aluminum, 
providing excellent throughput longward of 1000 \AA. Although the reflectivity 
of LiF drops substantially shortward of 1000 \AA, IMAPS achieved a useful 
throughput even in the 930 -- 980 \AA\ region.  The spectral resolution in IMAPS 
spectra obtained during the 1996 flight  was approximately $\lambda/\Delta\lambda \sim 80,000$, or $\le 4 \kms$.   The telluric \oi\ lines (e.g. \wl 950.112 near 
\lyd\ - see Fig \ref{gvlylines}) have FWHM $\sim 5 \kms$, but these lines are 
probably partly resolved in IMAPS spectra (see Jenkins et al. 1999b).

\gvel\ and \zpup\ were chosen for this IMAPS Guest Investigator program  because they were available in 1996 November, have a flux near \lyd\ $> 10^{-9} \ergcmsA$, have $v \sin i \ga 100 \kms$, and \nhi $\la 10^{20} \cmsq$.

\gvel\ was observed by IMAPS on 1996 November 27 14:34--15:11 and 16:00--16:47 
UT for a total of 4224 s. \zpup\ was observed on 1996 November 26 18:44--19:23 
and 20:16--20:57 UT also for a total of 4224 s. The exposure time at each 
echelle grating position ranged from 817.6 s to 1226.4 s.  These exposure times 
were chosen to obtain spectra with good signal-to-noise (S/N) ratios near the 
cores of the \di\ \lyd\ (949.485 \AA) and \lye\ (937.548 \AA) lines.  Typically, we found 
S/N=10--15 in the local continuum near the \lyd\ and \lye\ interstellar 
features.

\section{COLUMN DENSITY OF ATOMIC DEUTERIUM}\label{N(DI)}
\subsection{General Considerations}\label{x2method}

The data reduction and analysis of the \gvel\ and \zpup\ spectra and the 
determination of the \di\ column densities were identical to those described in 
\pap1\ for the line of sight toward \dori. IMAPS spectra were analyzed to determine the 
total \ndi\ (\S\ref{gvelN(DI)} and \S\ref{zpupN(DI)}). A large number of {\it 
International Ultraviolet Explorer} (\iue) high resolution \lya\ spectra were 
obtained from the National Space Science Data Center archive to determine the total \hi\ column density 
(\nhi) in \S\ref{N(HI)}. We note that given that \nhi\ $\sim10^{20} \cmsq$ 
on these sightlines and that D/H $\sim 10^{-5}$, the gas that gives rise to the 
\hi\ \lya\ damping wings is the same gas responsible for the \di\ \lyd\ and 
\lye\ features.

In \S\ref{NIprof} we show that \ni\ is a good tracer of \hi\  and \di\ 
on these sight lines. The line of sight column density per unit velocity [\nav , see \pap1] for \ni, defined by a range of \ni\ 
lines recorded in the IMAPS spectra, 
provided a velocity template for modelling the \di\ profiles.  The high S/N for 
the \ni\ lines helped to constrain the model profiles 
that gave an acceptable fit 
to the lower S/N \di\ lines. This can prevent noise in the \di\ profiles from giving 
arbitrarily large or small \ndi\ at specific velocities.  We did not decompose 
the velocity profiles of \ni\ or \di\ into separate (blended) Gaussian 
components -- the D/H ratios determined here were based on total column densities for each sight line.

\begin{figure}[h!]
\plotone{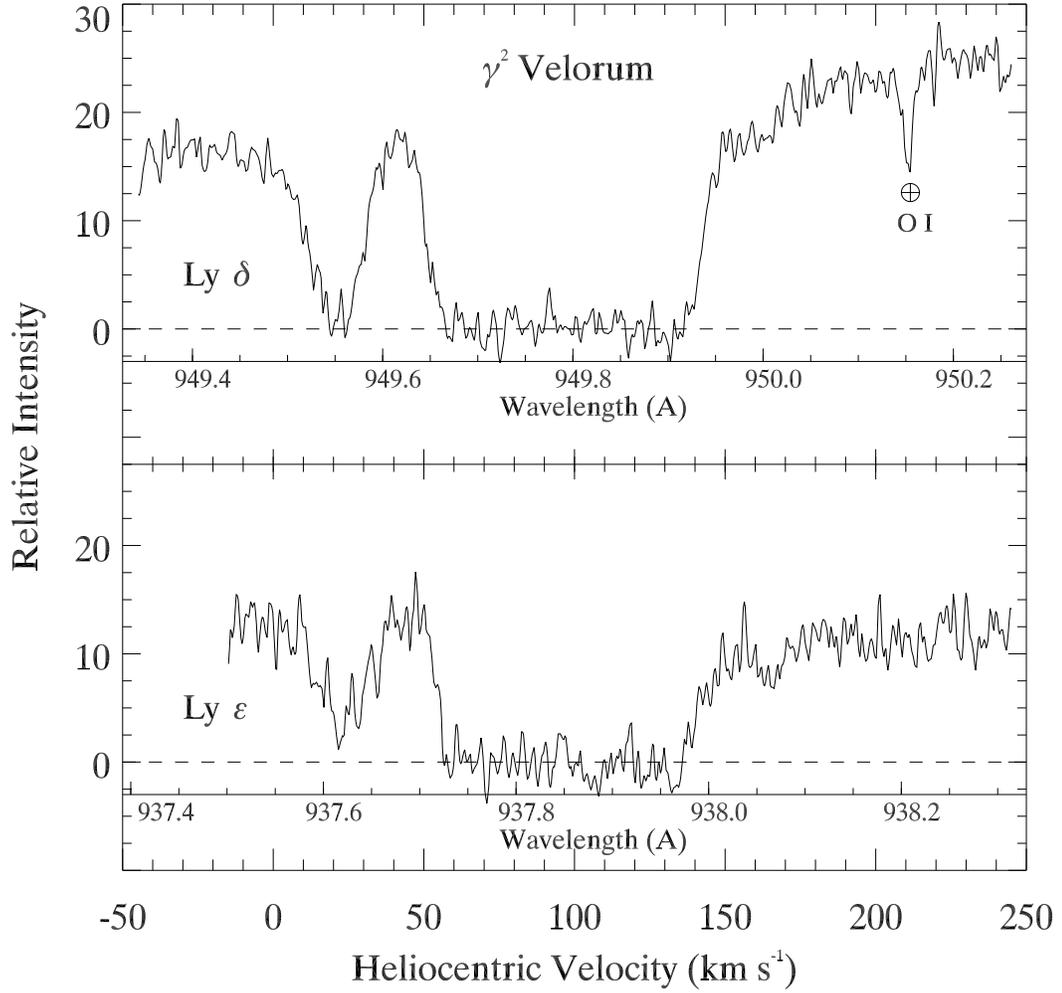}
\caption[]{Line profiles of \lyd\ and \lye\ from IMAPS spectra for 
\gvel. The zero-point of the velocity scale is computed with respect to the 
laboratory wavelengths of the \di\ lines. The narrow width of \oi * \wl 950.112 
telluric line indicates the spectral resolution is $\lambda/\Delta\lambda\sim 
80,000$. \label{gvlylines}}
\end{figure}

\begin{figure}
\plotfiddle{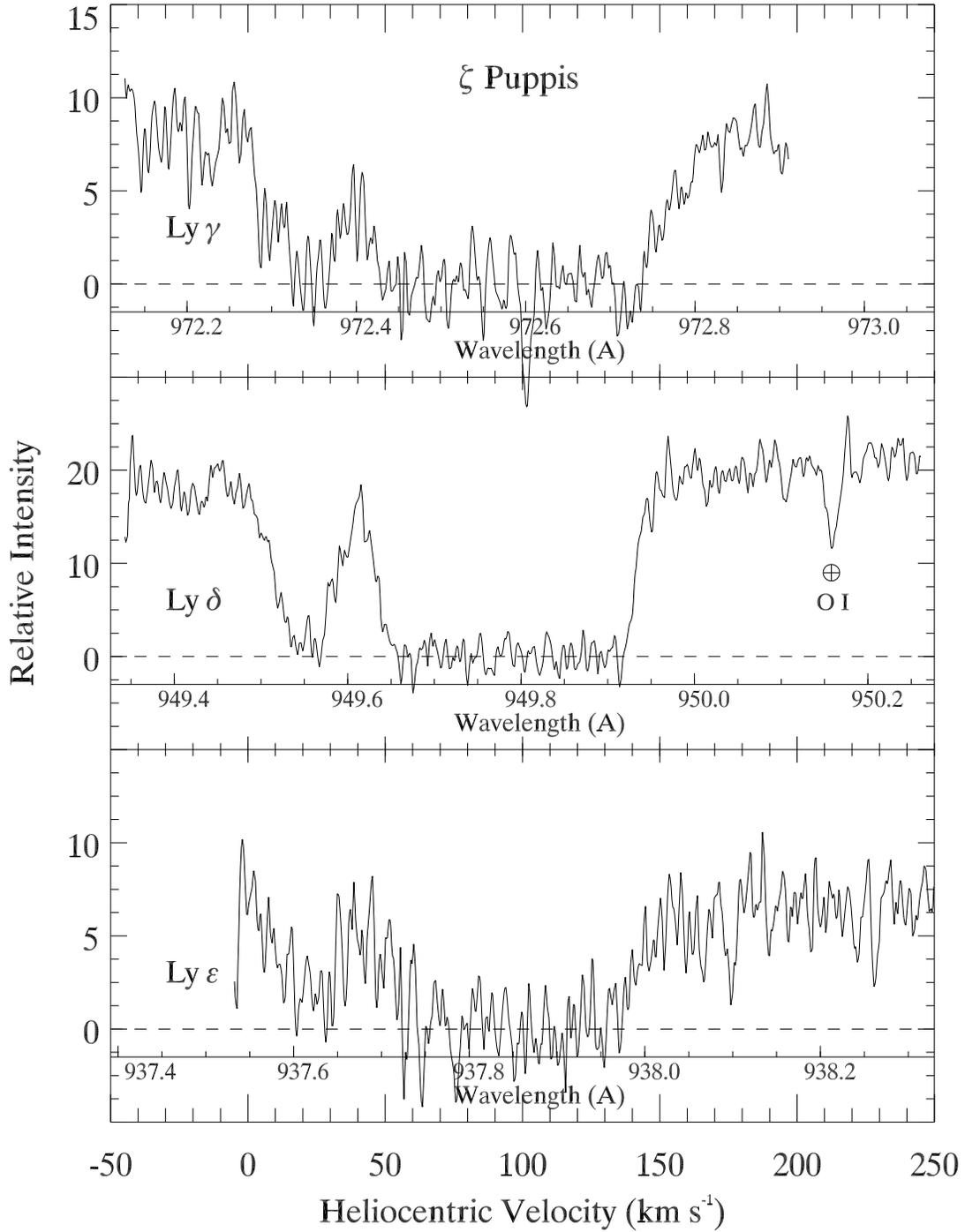}{19cm}{0}{90}{90}{-300}{-20}
\caption[]{Line profiles of \lyg, \lyd, and \lye\ from IMAPS spectra for 
\zpup. The zero-point of the velocity scale is computed with respect to the 
laboratory wavelengths of the \di\ lines.  \label{zplylines}}
\end{figure}

Using the method outlined in \pap1, we corrected  for the effects of the weak  
lines \feii\ \wl937.652  and \h2\ Lyman 14--0 $P$(2) 
\wl949.351.    The \feii\ line is located at $+51 \kms$ in \gvel\ and at $+53 \kms$ in \zpup\ on the heliocentric velocity scale whose zero point is the laboratory wavelength of \di\ \lye\ (see Figs. \ref{gvlylines} and \ref{zplylines}).  The \h2\ line is located at $+10 \kms$ in \gvel\ and at 
$+12 \kms$ in \zpup\ on the \di\ \lyd\ heliocentric velocity scale.
   The strength and shape of these features were 
computed from other transitions of the same species recorded at longer 
wavelengths (and higher S/N) in the IMAPS spectra of \gvel\ and \zpup.  The computed central residual intensities of these 
features are nearly identical for the two stars:  0.66 for \feii \wl937.6 and 0.95 for \h2 \wl949.3. The other 
potentially contaminating lines noted in Table 1 of \pap1\ can safely be 
ignored.

The possible contamination of the \lyd\ order by scattered light 
from  an adjacent order that contains the strong absorption lines of the
\ni\ \wll954 multiplet  was evaluated in the same 
manner as described in \pap1.  The \chisq\ analysis  found that any 
residual contamination of the spectrum in the vicinity of \lyd\  
by the pattern of saturated \ni\ features was $<1$\% of 
the continuum in both \gvel\ and \zpup. Amplitudes for this correction 
larger than $\sim1$\% caused unacceptably bad deviations in the bottom of the 
\hi\ \lyd\ profile.  Given the general noise characteristics 
of the \lyd\ profiles, this is a negligible effect.

For the case of \gvel, there is another potential contamination
source.  A companion star, $\gamma^1$~Vel, is located 42\arcsec\ to the
southwest of \gvel.  Its light should be accepted along with
that from \gvel, since IMAPS is an objective-grating
spectrograph without an entrance slit to reject unwanted sources.  (The
grid collimator rejects light coming from more than 1\arcdeg\ from the
axis, however.)  $\gamma^1$~Vel is 2.5 magnitudes fainter than
\gvel\ in $V$ and has a spectral classification of B1$\,$IV
(Hoffleit \& Jaschek, 1982).  Even though $\gamma^1$~Vel is cooler than
\gvel, its flux is not diminished much below the peak of the
Planck distribution at our wavelengths of interest, so we expect its
intensity at a given wavelength to be not more than about a factor of 10
fainter than \gvel.  Fortunately, when the observations were
taken the roll angle of the spacecraft about the optical axis of IMAPS
(governed by the position of the Sun in the sky) was such that the faint
spectrum of $\gamma^1$~Vel was displaced along the cross dispersion
direction toward the long-wavelength part of the format.  As a
consequence, any spectral segment in the spectrum of \gvel\ had
light superposed on it from shorter wavelengths in $\gamma^1$~Vel (the
separation was slightly less than the distance between 3 echelle
orders).  The rapid decline in the sensitivity of IMAPS toward shorter
wavelengths thus amplified the factor of 10 disparity in relative fluxes
at any given position on the format.  Therefore, the contamination of the \gvel\ spectrum 
by light from $\gamma^1$~Vel is negligible.  As a final note, on visual
inspection we are unable to see any ghost-like spectrum of
$\gamma^1$~Vel on top of the spectrum of \gvel.

The Lyman line profiles for \gvel\ and \zpup\ are shown in Figures 
\ref{gvlylines} and \ref{zplylines}.  These profiles  have 
been corrected for \feii\ and \h2\ line absorption, as described above.  
The background levels near the \di\ lines were determined from the 
broad, saturated cores of the adjacent \hi\ profiles.  As explained in \pap1, we compared resolution-degraded forms of the IMAPS profiles
with those recorded by \oao\ to test the proposition that the
cores of the \hi\ lines indeed represented the zero-intensity levels in
the vicinity of the adjacent \di\ features.  This was done to check
that we were not being misled by an effect from possible strong, broad
wings extending away from the main peak in the instrumental profile of
IMAPS. We concluded that indeed the cores of the \hi\ profiles provided 
very good estimates of the zero levels in the vicinity of the \di\ lines.

The best answers for \ndi\ and the deviations permitted by the data were 
evaluated by minimizing \chisq\ (see Lampton, Margon \& Bowyer 1976 and 
Bevington \& Robinson 1992 for details) when all of our unknown parameters were 
allowed to vary.  The ten free parameters for \gvel\ are (1) \ndi, (2) the gas 
temperature $T$, (3-8) the continuum slopes, Y-intercepts, and background levels for \lyd\ and \lye, (9) a shift in the velocity zero point between the 
\ion{D}{1} lines and the \ni\ template, and (10) a coefficient for scaling the 
\ni\ contamination signal in \lyd\ (see \S 3.2 in Paper I). The gas temperature is an 
important parameter because the \di\ line can be broadened significantly 
compared to the \ni\ template by thermal motions; see \S 4.1 in \pap1\ 
for details. There are three additional free parameters for \zpup\ because of 
the addition of \lyg\ to the analysis. Note that the linear continuum fitting (with specific velocity limits as given in \S\S\ref{gvelN(DI)} and 
\ref{zpupN(DI)}) is an integral part of the \chisq\ evaluation, i.e., the 
deviations of the continuum levels are considered, in addition to the behavior 
inside the \di\ profiles.  The same holds for the zero level 
as defined by the bottom of the adjacent \hi\ profile. We used Powell's method (Press et al. 1992, p. 406) to find the minimum \chisq.  We then set confidence limits by increasing (or decreasing) 
\ndi\ and $T$ with the other parameters freely varying until \chisq\ increased 
by the appropriate amount for the confidence limit of interest with two useful 
parameters, \ndi\ and $T$. We also used different initial values to establish 
that the minimum \chisq\ is unique.
\clearpage
\placefigure{gvlylines}

\placefigure{zplylines}
\clearpage

\subsection{Velocity Profile Templates}\label{NIprof}

In \pap1\ we discussed the benefits of obtaining a velocity profile 
based on high quality data for \ni\ and \oi, two species that are not 
significantly depleted in the ISM and  which have very similar 
ionization balances to those of \di\ and \hi\ (Ferlet 1981; York et al. 1983).  
This profile information is 
helpful in constraining the variety of possible interpretations that 
would produce acceptable fits with the D profiles.  For \ni, we used 
IMAPS spectra of the 10 lines in the multiplets at 952.4, 
953.8, 954.1 and 1134.7 \AA \footnote{One of the lines in the 1134.7 \AA\ 
multiplet, the line at 1134.165 \AA , had to be 
omitted from consideration for \zpup\ because it was too close to the 
edge of the image format.}. 
For all of the \ni\ lines except those in the 952.4 \AA\ multiplet, the 
background level was easily defined because the stronger lines were saturated.  
Of course, these lines were useful only for defining the behavior of 
the gas at velocities somewhat removed from the line core.  Since the 
lines in the multiplet at 952.4 \AA\ were not saturated, we had to 
determine the background level by a different method.  For both \gvel\ 
and \zpup, there are U1 scans of this weak multiplet available in the 
archive of spectra recorded by the \oao\ satellite (Rogerson et al. 1973).
After comparing our IMAPS spectrum degraded to the resolution of 
\oao\ with the actual \oao\ scans, we determined 
the adjustments to the background levels that were needed for the IMAPS 
spectra of this \ni\ multiplet.  The backgrounds caused by grating 
scatter in the  
\oao\ spectra were taken from the observed count rates in the 
bottoms of the hydrogen \lyg\ and \lyd\ lines, and these levels were 
subtracted off before the comparison was made.

As was done for our analysis in \pap1, we adopted a method developed by Jenkins 
\& Peimbert (1997) to create a composite profile for the column density of \ni\ 
as a function of velocity from the 10 lines in the four multiplets, with the 
weak lines in each case defining the main part of the profile and the strong 
lines outlining the exact shape of the profile's wings, well away from the 
saturated part of the line.  For places where there was overlap in the useful 
portions of the lines, there was satisfactory agreement.  The $f$-values of 
Goldbach et al. (1992) were adopted for our analysis.

\begin{figure}
\plotfiddle{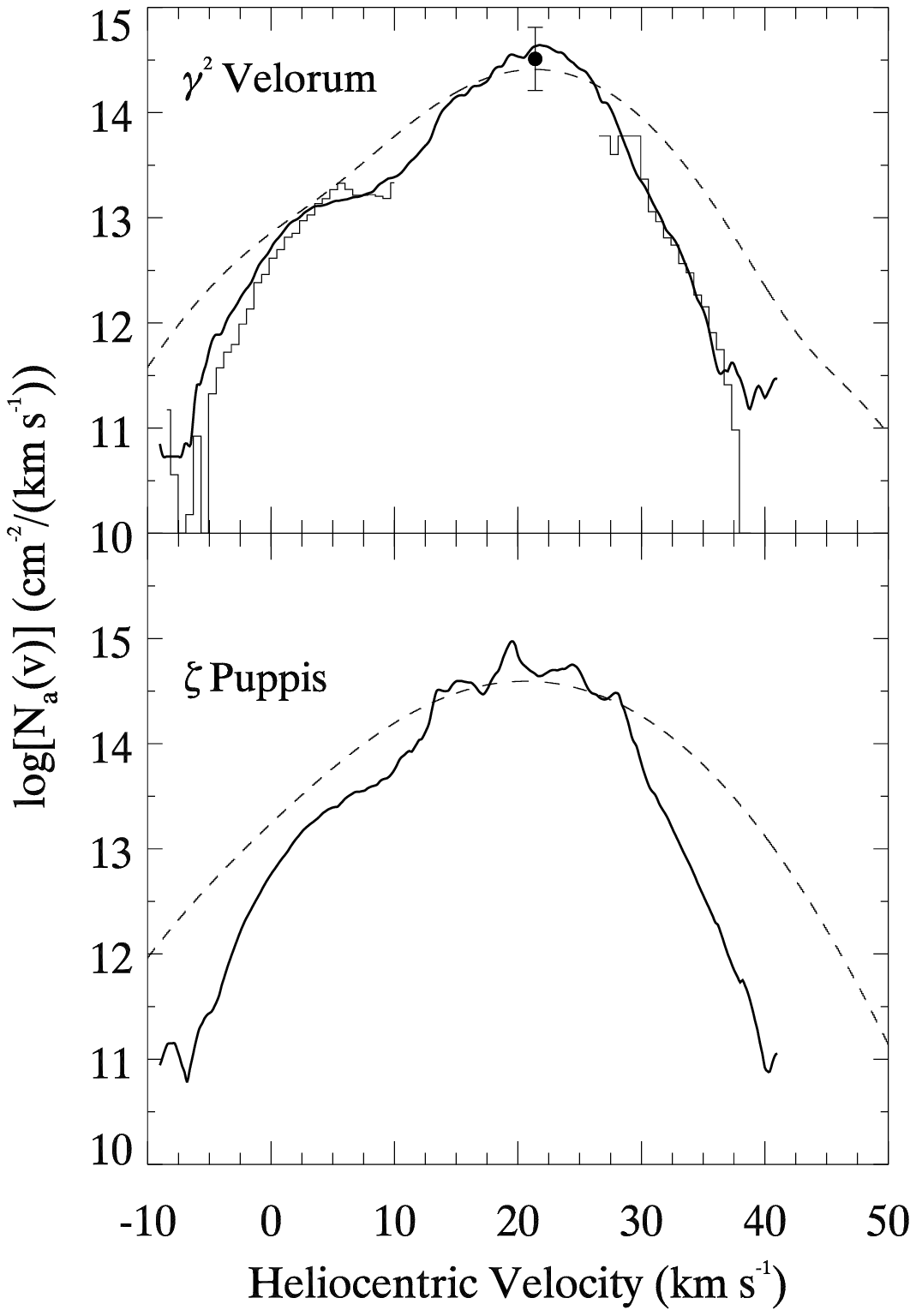}{15cm}{0}{100}{100}{-220}{-230}
%\plotfiddle{fig3_niplots2.ps}{18cm}{0}{100}{100}{-240}{-240}
\caption[]{Profiles for \nav\ (thick solid lines) for \ni\ toward  \gvel\ and  
\zpup\ derived from IMAPS spectra. For \gvel, \nav\ for the wings of \oi\ \wl 
1039.23, adjusted for the difference in cosmic abundance between O and N 
($-0.90$ dex), is shown by the thin histogram line. 
The single data point and error bar 
correspond to $N$(\oi) from Fitzpatrick \& Spitzer (1994) for \oi\ \wl 1355.6 
(see text for details). The dashed lines show the expected shapes of the \di\ 
profiles for a thermal broadening for $T=6070$ K (\gvel) and $T=9550$ K (\zpup) of the respective \ni\ profile favored by our most likely solutions in Tables 
\ref{gvdtab} and \ref{zpdtab}. \label{NI_Navfig}}
\end{figure}

Figure \ref{NI_Navfig} shows the \nav\ profiles derived for \ni\ 
toward \gvel\ and \zpup.  In \pap1\  we showed that the profiles are unlikely to be contaminated by additional contributions from telluric absorption.  \nni\ is $\sim$1\% of \noi\ in the Earth's atmosphere at the altitude of IMAPS ($\sim$305 
km) during the \orf\ mission. Even for our strongest \ni\ transition used to derive the far wings of \nav\ for \ni\ (\wl1134.98), the telluric absorption should be only one third as strong as the \oi * \wl950.112 feature present in Figures \ref{gvlylines} and \ref{zplylines}, and at a heliocentric velocity of $+12.6 \kms$ it would be buried in the saturated portion of the profile. Consequently, telluric \ni\ has a negligible effect on the 
interstellar \ni\ profiles.

\placefigure{NI_Navfig}

\oi\ is the best tracer of \hi\ in the ISM since the ionization potential of 
\oi\ (13.56 eV) is nearly identical to that of \hi.  The ionization potential of \ni\ (14.53 eV) is 
only slightly greater that that of \hi. Furthermore, both O and N are coupled to H by resonant charge exchange reactions. As discussed by Sofia \& Jenkins (1998) 
and Jenkins et al. (2000), \ni\ closely follows \oi\ and \hi\ unless $n_{\rm e} 
\gg \ n_{\rm H\,I}$. 
 
In the IMAPS wavelength band there is no suitable set of \oi\ transitions to 
completely define a velocity profile template. The available \oi\ 
lines are highly 
saturated or are not detectable, either because they are buried in the core of 
a Lyman series line or because they are too weak.  In \pap1\ we made use of an 
archival \hst\ spectrum of the very weak intersystem transition of \oi\ at 
1355.6 \AA\ for \dori.  There is no such spectrum available for \zpup, and the 
measurement of this line in the spectrum of \gvel\ taken by  Fitzpatrick \& 
Spitzer (1994) is too noisy to adequately define the shape of the \oi\ profile. 
Therefore, we were forced to use the next best option, \ni, to define the velocity profile template.

Although we do not have a complete \nav\ profile for \oi, we tested the 
assumption that \ni\ traces \hi\ by comparing  selected portions of the \oi\ \nav\ profile derived from available line profiles in \gvel.  
First, we computed \nav\ from the wings of \oi\ \wl 1039.230. Second, we used 
the GHRS spectrum of \oi\ \wl 1355.6 (Fitzpatrick \& Spitzer 1994) to define
\nav\ in the core of the line. The computed \nav\ was scaled 
by the difference in the N and O 
solar abundances ($-0.90$ dex, Anders \& Grevesse 1989). The results, shown in 
Fig \ref{NI_Navfig}, demonstrate that the wings and core of \nav\ for \ni\ and 
\oi\ are in excellent agreement.  Consequently, we have high confidence that
the \nav\ profile for \ni\ provides an accurate model for the velocity distribution of \hi.

\subsection{\gvel orum}\label{gvelN(DI)}

The continuum near the \lyd\ line shown in Figure \ref{gvlylines} was determined 
in the velocity range $(-40, -10)\kms$ on the blue side of the \di\ line and 
$(+170, +205)\kms$ on the red side. For \lye, the continuum limits were $(-14, 
+8)\kms$ and $(+170, +210)\kms$. The background level was determined from the 
core of the two \hi\ lines in the velocity range $(+60, +140)\kms$. The \di\ 
profiles were fit in the velocity range $(-10, +39)\kms$ for \lyd\ and $(+8, 
+39)\kms$ for \lye.  With a spacing of independent velocity samples of 
$1.25\kms$, this resulted in 276 degrees of freedom in the \chisq\ analysis to 
fit the 10 free parameters described in \S\ref{x2method}. 

\begin{figure}
\plotone{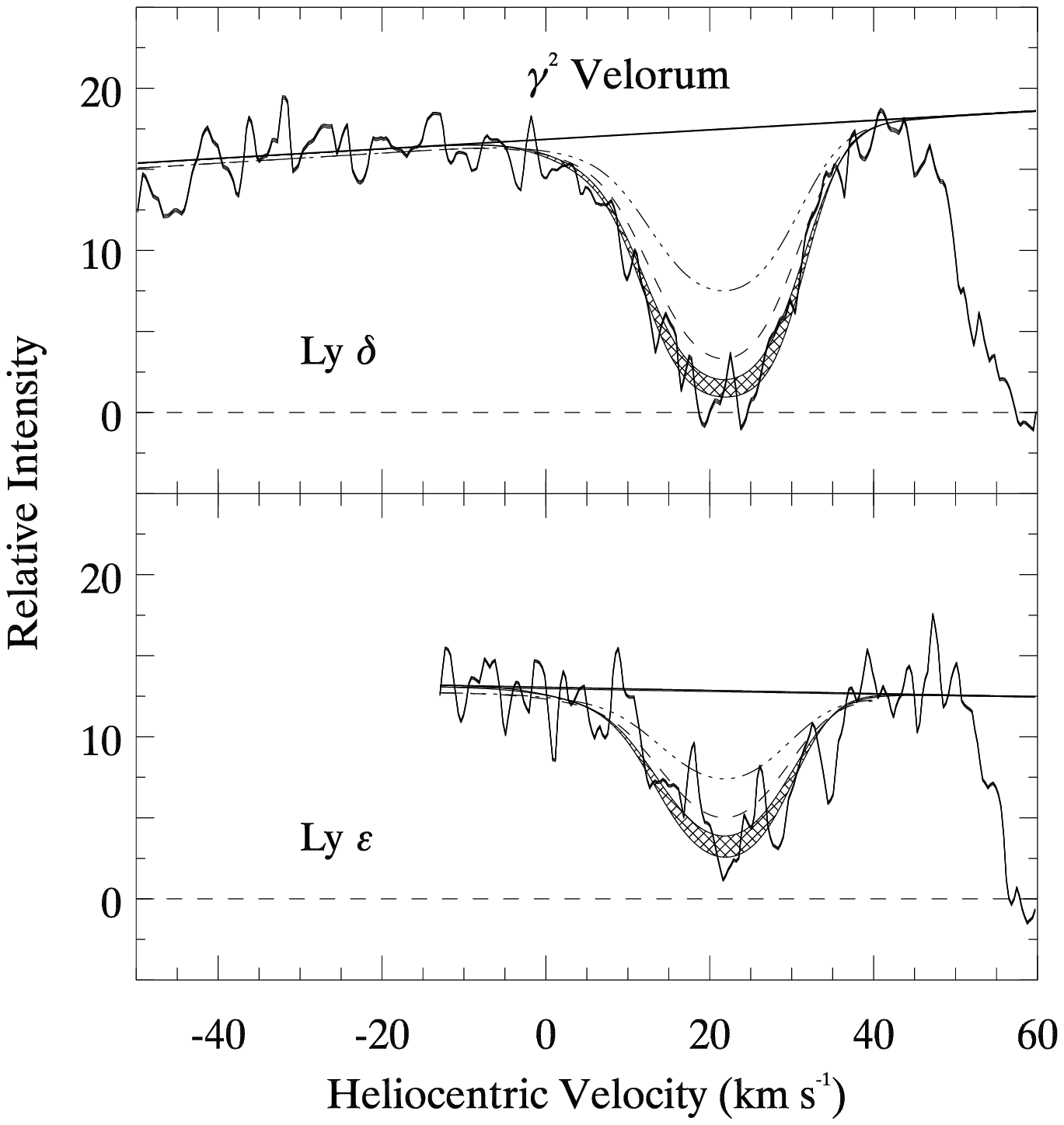}
\caption[]{Observed and model line profiles of \di\ \lyd\ and \lye\  for 
\gvel\ showing the region bounded by the 90\% confidence limits on \ndi\ (cross-
hatched region), derived from both lines collectively. Assuming the value of 
\nhi\ derived for \gvel\ in \S\ref{N(HI)}, the dashed line illustrates the \di\ 
line shapes necessary to achieve D/H $=1.5\EE{-5}$ and the dot-dashed line 
corresponds to the \di\ profiles necessary to obtain D/H $=0.74\EE{-5}$ (\dori, 
\pap1). \label{gvdfits}}
\end{figure}

By adjusting our estimate for the noise, which was only known to limited accuracy beforehand, we achieved a minimum value for \chisq\ equal to 247.0 at a 
column density of $1.12\EE{15}\cmsq$.  For 276 degrees of freedom, there was 
a 90\% chance that we would have found this value of \chisq\ or greater.  
With this 90\% 
confidence level, we arrived at a conservatively high estimate for the noise.  It then followed that this noise level, 
which is perhaps higher than the real noise, 
gave a conservatively large confidence interval for \ndi, as determined by how 
rapidly the values of
\chisq\ deviated from the minimum value.  With the noise level having
been set in the manner just described, we explored for 90\% and 99\%
confidence limits for \ndi\ (i.e., $1.65\sigma$ and 2.58$\sigma$ deviations),
which correspond to \chisq(min)+4.6 and \chisq(min)+9.2, respectively.  Table 
\ref{gvdtab} lists \ndi\ and $T$ for the best value and these limits.  The 
$\pm$90\% confidence limits on the model  \di\ profiles ({\it cross-hatched 
regions}) for \lyd\ and \lye\ are compared with the observed deuterium profiles 
in Figure \ref{gvdfits}.

\placetable{gvdtab}
\begin{deluxetable}{cccc}
\tablewidth{0pc}
\tablecaption{Limits for \ndi Toward \gvel\label{gvdtab}}
\tablehead{~ & \ndi\ & $T$ &  \nl
Significance & $(10^{15} \cmsq)$ & (K) & \chisq 
 \ }
\startdata
Minimum \ndi at the 99\% confidence limit \dotfill & 0.96 & 6630 & 255.6 \nl
Minimum \ndi at the 90\% confidence limit \dotfill & 1.00 & 6530 & 251.6 \nl
Best \ndi \dotfill & 1.12 & 6070 & 247.0 \nl
Maximum \ndi at the 90\% confidence limit \dotfill & 1.27 & 5280 & 251.7 \nl
Maximum \ndi at the 99\% confidence limit \dotfill & 1.34 & 4870 & 256.1 \nl
\enddata
\end{deluxetable}

\placetable{zpdtab}
\begin{deluxetable}{cccc}
\tablewidth{0pc}
\tablecaption{Limits for \ndi Toward \zpup\label{zpdtab}}
\tablehead{~ & \ndi\ & $T$ &  \nl
Significance & $(10^{15} \cmsq)$ & (K) & \chisq 
 \ }
\startdata
Minimum \ndi at the 99\% confidence limit \dotfill & 1.07 & 10600 & 397.7 \nl
Minimum \ndi at the 90\% confidence limit \dotfill & 1.13 & 10400 & 393.1 \nl
Best \ndi \dotfill & 1.30 & 9550 & 388.5 \nl
Maximum \ndi at the 90\% confidence limit \dotfill & 1.49 & 8500 & 393.1 \nl
Maximum \ndi at the 99\% confidence limit \dotfill & 1.58 & 7900 & 397.7 \nl
\enddata
\end{deluxetable}

\placefigure{gvdfits}

In Figure \ref{gvdfits} we also show with a {\it dashed line} the expected shape 
and depth of the \di\ profile for \ndi=$7.7\EE{14} \cmsq$. This corresponds to a 
D/H ratio of $1.5\EE{-5}$, assuming the \hi\ column density derived below in 
\S\ref{N(HI)}. For this case, \chisq -- \chisq(min) = 52.4, which is clearly an 
unacceptable fit. The {\it dot-dashed line} in Figure \ref{gvdfits} corresponds 
to the \di\ profile for \ndi=$3.8\EE{14} \cmsq$, the \di\ column density 
necessary to achieve D/H=$0.74\EE{-5}$, the value found toward \dori\ in \pap1.
A similar demonstration for the measurement of \nhi\ is be given in \S\ref{D/H}.

\subsection{\zpup pis}\label{zpupN(DI)}

\begin{figure}
\plotfiddle{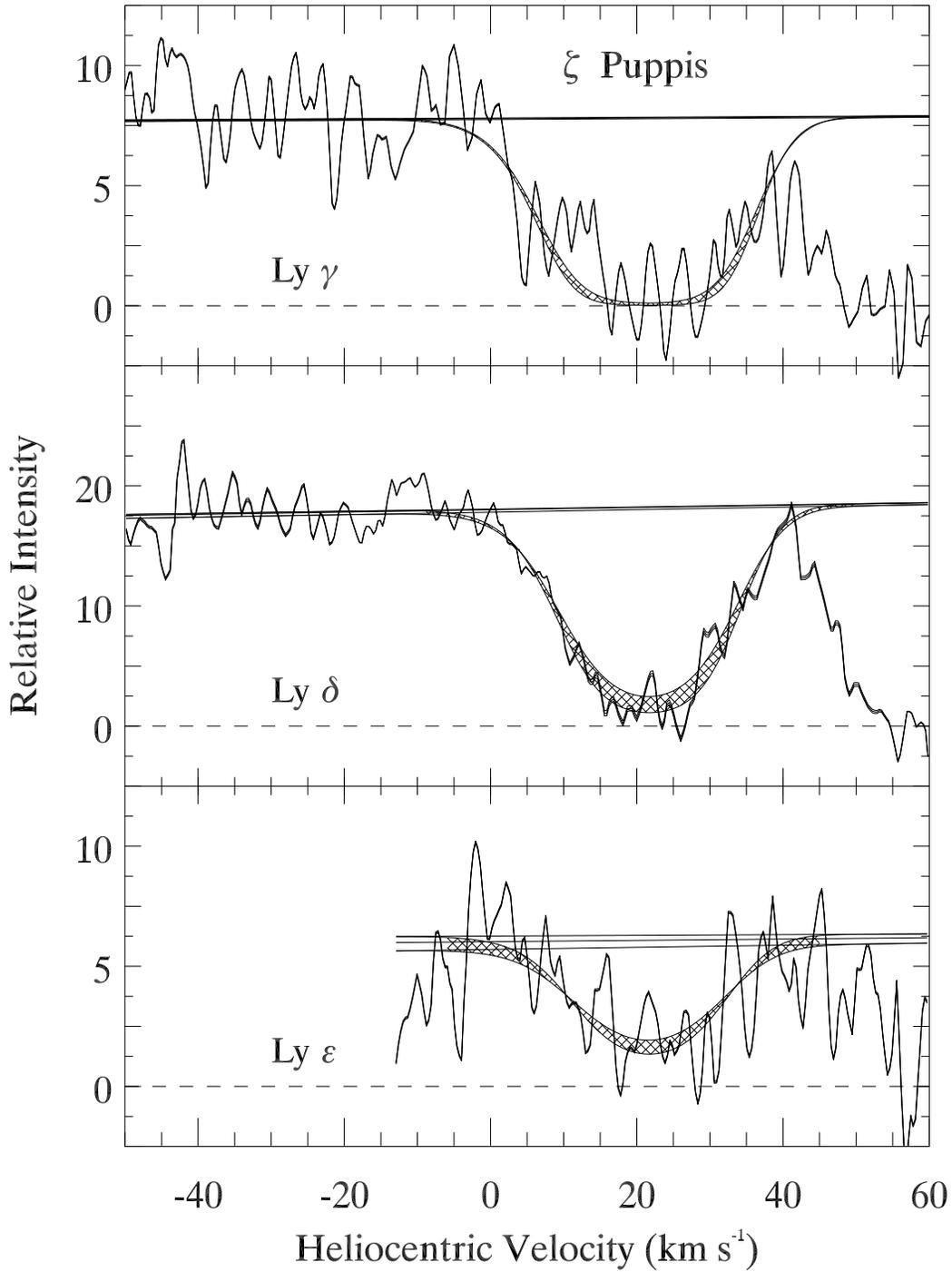}{19cm}{0}{100}{100}{-300}{-50}
\caption[]{Observed and model line profiles of \di\ \lyg, \lyd, and \lye\  for 
\zpup\ showing the region bounded by the 90\% confidence limits on \ndi\ (cross-hatched region), derived from all three lines collectively. 
Note that the preferred continuum levels near \lye\ change for 
different values of \ndi. \label{zpdfits}}
\end{figure}

The measurement of \ndi\ toward \zpup\ used the same methodology 
described in the previous section with the following modifications. In addition 
to \lyd\ and \lye\ we were able to include the \di\ profile of \lyg\ in the 
\chisq\ analysis. The blue and red regions for the linear continuum fits were  
$(-40, -10)\kms$ and $(+170, +190)\kms$ for \lyg, $(-39, -10)\kms$ and $(+157, 
+190)\kms$ for \lyd, and $(-4, +3)\kms$ and $(+180, +220)\kms$ for \lye. The 
background levels were determined from the saturated cores of the \hi\ lines 
over these velocity limits: $(+55, +130)$ for \lyg\ and \lyd, $(+65, +115)$ for 
\lye. The model \di\ profiles were fit over the range $(-5, +40)\kms$ for \lyg, 
$(0,+40)\kms$ for \lyd, and $(+3, +37)\kms$ for \lye. This resulted in 426 
degrees of freedom to simultaneously fit the 13 free parameters described in 
\S\ref{x2method}.  The minimum \chisq\ is 388.5 and corresponds to the most 
probable value of \ndi\ for \zpup, namely $1.30\EE{15}\cmsq$. The results of the 
\chisq\ analysis are summarized in Table \ref{zpdtab}, including the 90\% and 
99\% confidence limits.  The same conservative treatment of the noise described 
for \gvel\ was followed for \zpup.  Figure \ref{zpdfits} shows the observed \di\ 
Lyman line profiles and the range in the best fit model profiles and continua 
fits allowed by the 90\% confidence limits.

\placefigure{zpdfits}
\clearpage
\section{COLUMN DENSITY OF ATOMIC HYDROGEN}\label{N(HI)}
\subsection{Analysis Methodology}

The primary objective of this IMAPS program is to determine D/H ratios of 
sufficient accuracy to test for spatial inhomogeneities. To achieve this goal, 
we must strive for a precision 
in \nhi\ that is as good as (or preferably better than) that 
of the deuterium measurement. We must also understand 
the magnitude and nature of the uncertainties in \nhi\ and 
\ndi\ so that we can realistically assess the uncertainty in 
D/H for comparison to other measurements in the ISM. For these reasons, 
we decided to conduct our own investigation of the \hi\ column 
densities toward \gvel\ and \zpup, even though a number 
of measurements of \nhi\ toward these stars are already 
available in the literature (e.g. Jenkins 1971; York \& Rogerson 1976;
Vidal-Madjar et al. 1977; Bohlin, Savage, \& Drake 1978; Shull \& Van 
Steenberg 1985; Diplas \& Savage 1994).

We described our method for determining \nhi\ in \pap1; here we 
provide a summary. In \S\ref{zpupN(HI)} and \S\ref{gvelN(HI)} we discuss details 
particular to the \zpup\ and \gvel\ sight lines and present the results. We 
used the \lya\ line to determine the total \nhi.\footnote{The \oao\ studies of D/H on the \gvel\ and \zpup\ sight lines 
used the much weaker damping wings of \lyb\ or \lyg\ to determine \nhi.}
Since \hi\ \lyg\ and \lyd\ are on the flat part of the curve of growth, high velocity gas could influence  their profiles.  \lya\ is immune from this potential problem. Due to the great breadth of the Lorentzian wings ($\pm1000\kms$ -- see the 
velocity scale in Figure \ref{gvHIblowup}), \lya\ absorption due to any high 
velocity interstellar gas ($\pm100\kms$) that may be present is confined to the 
saturated core of the profile. In principle, some high velocity \hi, which is too widely dispersed in velocity to show up in the \di\ profiles, could contribute 
to the \lya\ damping wings. However, this gas is not detected in the strong 
lines  \ni\ \wl1134.98 and \oi\ \wl1039.23 observed with IMAPS.
It is clear that such high velocity gas, if present, would have a low column density and 
a negligible effect on the \lya\ damping wings.  Since the \lya\ line
is not covered by IMAPS, was not observed by the \hst\ spectrographs, and the available \oao\ \lya\ 
scans suffer from  several problems (see \pap1), we analyzed 
high dispersion \iue\ observations of \lya\ to determine \nhi.  
Furthermore, these 
stars were observed many times with \iue\ over the course of many  years, 
and this offers an opportunity to evaluate potential sources of  systematic 
error.  For example, \gvel\ is a spectroscopic  binary, and the large \iue\ 
database allowed a search for systematic changes in the derived \nhi\ as a 
function of orbital phase.

After screening and retrieving the \iue\ data\footnote{We used the 
standard IUESIPS calibration of the \iue\ high dispersion spectra instead of the 
NEWSIPS calibration to avoid potential problems with background corrections and 
zero-level determination near \lya\ found in some \iue\ SWP high dispersion 
spectra processed with NEWSIPS (see Massa et al. 1998), which could adversely 
affect our measurements.} 
of interest and correcting for interorder scattered light (see \pap1), we used 
an approach first used by Jenkins\markcite{j71} (1971) to constrain the \hi\ 
column density, i.e., we determined the \nhi\ which provides the best fit to the 
\lya\ profile with the optical depth $\tau$ at a given wavelength $\lambda$ 
calculated from the expression
\begin{equation}
\tau (\lambda ) = N({\rm H~I})\sigma (\lambda) = 4.26\times 10^{-20} 
N({\rm H~I})(\lambda - \lambda _{0})^{-2}
\end{equation}
where $\lambda _{0}$ is the \lya\ line center at the velocity 
centroid of the hydrogen (note that the useful portion of the \lya\ 
profile is entirely due to the Lorentzian wings and the effects of 
instrumental and Doppler broadening can be neglected). As was done for 
\di, we also determined the important free parameters which can 
be adjusted to fit the \hi\ \lya\ absorption profile, then we 
found the set of parameters that minimized \chisq\ using Powell's 
method. We set confidence limits on the \hi\ column as described in \S 3.1 with 
only one parameter of interest, \nhi . The other (uninteresting) free 
parameters we selected for fitting the \hi\ profile were three coefficients 
which specify a second-order polynomial fit to the continuum, and an 
additive correction to the flux zero point. The continuum was fit to the spectrum in several windows covering the range 1185 \AA\ to 1276 \AA. Despite our use of the Bianchi \& 
Bohlin (1984) correction for interorder scattered light, 
in many cases inspection of the flat-bottomed, saturated portion of the 
\lya\ profile showed that the zero intensity level was not quite 
correct. A zero point shift in the flux scale, one of the terms for evaluating \chisq, was determined from
a region within the saturated core.  For both stars, the zero point of the \lya\ wavelength scale was set by 
placing the \ni\ \wl1200 multiplet in agreement with the IMAPS \ni\ profiles. 
The \lya\ profile was then fit  only to the red wing because of the presence 
of strong stellar features superposed on the blue wing. However, the 
uncontaminated portion of the blue wing of \lya\ was checked for consistency 
with the fit to the red wing and found to be in good agreement.

\subsection{$\zeta$ Puppis}\label{zpupN(HI)}

In \pap1, considerable attention was paid to the spectroscopic binary nature 
of \dori\ and whether or not this could cause systematic 
errors in \nhi. A similar analysis of \gvel\ is presented in 
\S \ref{gvelN(HI)}. The determination of \nhi\ 
toward \zpup\ was less complex than 
the other stars observed by IMAPS to study D/H. As far as is known, 
\zpup\ is not a spectroscopic binary star. It is an O4 supergiant, 
so the stellar \hi\ \lya\ absorption line makes a negligible 
contribution to the \hi\ absorption profile. The star is 
possibly a non-radial pulsator with very weakly variable stellar 
absorption lines with a period of 8.5 hr (Reid \& 
Howarth\markcite{rh96} 1996), and the variability of the stellar wind 
P-Cygni profiles is well known and studied (e.g., Prinja et 
al.\markcite{p92} 1992; Howarth, Prinja, \& Massa\markcite{hpm95} 1995) 
and shows significant power at 19.2 hr and 5.2 day periods. Both of 
these sources of spectral variations are expected to have little impact 
on the interstellar \hi\ column density derived from \lya, 
but this can be checked given the large number of observations and good 
temporal sampling.

There are more than 200 high-dispersion observations of \zpup\ in 
the \iue\ archive, primarily because the star was selected for
intensive \iue\ observing programs to study stellar wind variability in massive 
stars. We  concentrated on the spectra obtained in three specific multi-day 
observing runs in 1986, 1989, and 1995. In 1995 \zpup\ was 
observed continuously for 16 days (dubbed the ''MEGA`` campaign, Massa et 
al.\markcite{m95} 1995).  We omitted three observations from these observing 
sessions because the archival data were corrupted or unavailable. This left 189 
observations for measurement of \nhi. All of the 
observations were obtained with the \iue\ large aperture and the 
signal-to-noise ratios are comparable; the \zpup\ data set is more uniform 
than the \iue\ \lya\ data used for measuring \nhi\ toward \dori\ and \gvel.

\begin{figure}
\plotfiddle{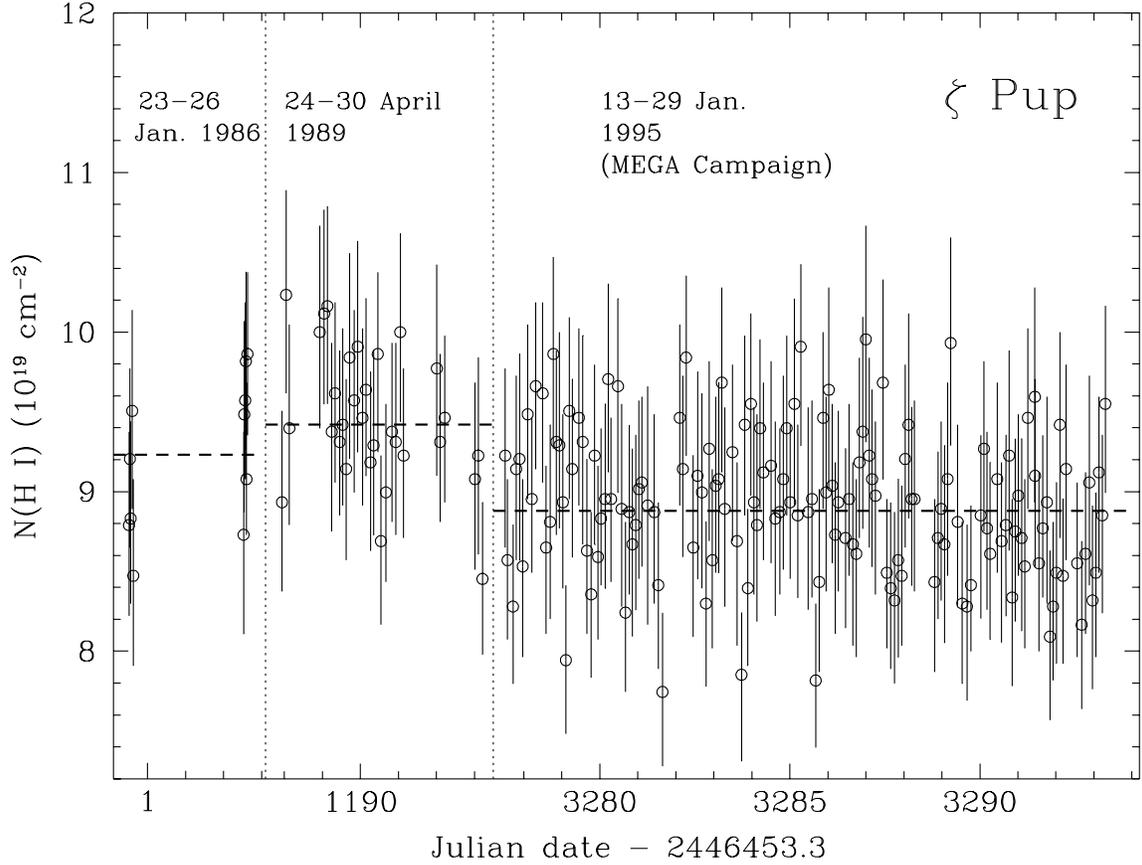}{12cm}{270}{60}{60}{-230}{390}
\caption[]{\hi\ column densities toward \zpup\ derived from 
the 189 high-dispersion \iue\ observations obtained in 1986, 1989, 
and 1995, plotted versus the observation date. Each measurement is 
shown with $\pm 1\sigma$ error bars. All observations were obtained 
with the large aperture and have comparable signal-to-noise ratios. 
Each data set is separated by a vertical dotted line (i.e., there is a 
break in the x-axis at each vertical dotted line), and the mean 
\hi column densities derived from the three data sets are 
indicated with heavy dashed lines. There appears to be a systematic 
difference between the column densities derived from the 1995 data and 
the earlier observations.\label{zpHIfig}}
\end{figure}

\placefigure{zpHIfig}

\placetable{zpnhi}
\begin{deluxetable}{lccccc}
\tablewidth{0pc}
\tablecaption{Interstellar \nhi\ Toward \zpup\label{zpnhi}}
\tablehead{Data & No. & $<N$(H~I)$>$ & $\sigma$\tablenotemark{b} & 
$\epsilon$\tablenotemark{c} & $\chi^{2}_{\nu}$\tablenotemark{~d} \nl
Set\tablenotemark{a} & Spectra & ($\cmsq$) & ($\cmsq$) & 
($\cmsq$) & \ }
\startdata
1986 & 11 & 9.23$\times 10^{19}$ & 0.47$\times 10^{19}$ & 0.14$\times 
10^{19}$ & 0.69 \nl
1989 & 31 & 9.42$\times 10^{19}$ & 0.43$\times 10^{19}$ & 0.08$\times 
10^{19}$ & 0.56 \nl
1995\tablenotemark{e} & 147 & 8.88$\times 10^{19}$ & 0.46$\times 
10^{19}$ & 0.04$\times 10^{19}$ & 0.69 \nl
Adopted value\tablenotemark{f} & 189 & 9.18$\times 10^{19}$ &  \ldots
 & \ldots & \ldots 
\enddata
\tablenotetext{a}{All \nhi\ measurements were derived from \iue\ 
high dispersion spectra using the large aperture. We have used only
the observations obtained during three observing runs in 1986, 1989, and 1995. 
This table shows the mean \hi\ column 
density, $<$\nhi$>$, derived from the 1986 data only, the 1989 data 
only, and the 1995 data only.}
\tablenotetext{b}{$\sigma$ = rms dispersion (both $\sigma$ and 
$<$\nhi$>$ were weighted inversely by the variances of the individual 
\nhi\ measurements).}
\tablenotetext{c}{$\epsilon$ = error in the mean = $\sigma$/(No. 
measurements)$^{0.5}$.}
\tablenotetext{d}{Reduced $\chi ^{2}_{\nu} = \chi ^{2}$/(degrees of 
freedom), where $\chi ^{2}=\sum_i \left\{ \left[ N_i({\rm H~I})- <N({\rm 
H~I})>\right]/\sigma\left[ N({\rm H~I})\right]_i \right\}^2$.}
\tablenotetext{e}{The \iue\ MEGA campaign (Massa et al. 1995).}
\tablenotetext{f}{Final adopted value of \nhi\ is the average of $<$\nhi$>$ for 
the three data sets. See text for details.}
\end{deluxetable}

The \nhi\ values derived from each individual observation are 
plotted in Figure \ref{zpHIfig}, along with the 1$\sigma$ 
uncertainties, versus the observation date. Table \ref{zpnhi} 
summarizes the mean \hi\ column density $<$\nhi$>$ and 
the root mean square (rms) dispersion $\sigma$ obtained from the 1986, 
1989, and 1995 data sets analyzed separately, and $<$\nhi$>$ 
for each data set is indicated with a heavy dashed line in Figure 
\ref{zpHIfig}. We also list in Table \ref{zpnhi} the formal 
error in the mean $\epsilon = \sigma /\sqrt{N} $ (here $N$ is the 
number of measurements) and the reduced \chisq\ for the three data 
sets. 

In \pap1\ we discussed the potential sources of 
systematic error in the derivation of \nhi\ from \iue\ 
spectra, and we suggested that the real uncertainty in the mean is likely to 
be greater than the formal estimate. The very large \lya\ data set for \zpup\ 
shown in Figure \ref{zpHIfig} appears to confirm that this is indeed the case
for this star. 
From this figure one can see that the column densities derived from the 1995 
data are systematically lower than the column densities derived from the 1989 
observing run. The mean of the 1995 data set is lower than the mean of the 
1989 data by $0.54\EE{19}\cmsq$ (0.0256 dex), and this difference is 
substantially greater than the formal error estimates ($\epsilon$). 

This systematic error could be the result of secular changes in the stellar mass 
loss or circumstellar environment, a previously unrecognized stellar variability 
with a long period, or instrumental effects. Or perhaps some aspect of the 
interstellar sight line changed during this period. Since the IMAPS observations 
for \ndi\ were made almost two years after the MEGA campaign, it is possible 
that a similar systematic error is present in any \iue\ estimate of \nhi\ that 
we choose for comparison  with the IMAPS \ndi. The exact magnitude of this 
systematic uncertainty is difficult to estimate, but we find that \nhi\ obtained 
from additional large aperture spectra of \zpup\ taken at other epochs are in 
good agreement with the estimates in Table \ref{zpnhi}. This 
indicates that the systematic error is not much larger than 0.0256 dex. 
For example, large aperture observations taken in 1988 May, 1989 December, and 
1991 March, combined with the three $<$\nhi$>$ values for 
the three data sets in Table \ref{zpnhi}, yield a mean 
\nhi\ = $9.34\pm0.67\EE{19}\cmsq$.  We conclude that 0.0256 dex 
is a conservative estimate of the $1\sigma$ uncertainty in \nhi. We therefore 
adopt the unweighted mean of the three $<$\nhi$>$ values in Table \ref{zpnhi} as the best estimate of \nhi\ toward \zpup\ and 0.0256 dex as its $1\sigma$ 
uncertainty: \nhi$=9.18\pm0.54\EE{19}\cmsq$.
We elected to use a mean of the $<$\nhi$>$ values in
Table \ref{zpnhi} instead of a straight mean of
all of the individual measurements shown in Fig. \ref{zpHIfig}
because the latter would give excessive weight to the
1995 data set due to the much larger number of measurements
obtained during that observing campaign.

\subsection{\gvel orum}\label{gvelN(HI)}

\begin{figure}
%\plotfiddle{fig7_gvhi1.ps}{12cm}{90}{68}{68}{270}{0}
\plotfiddle{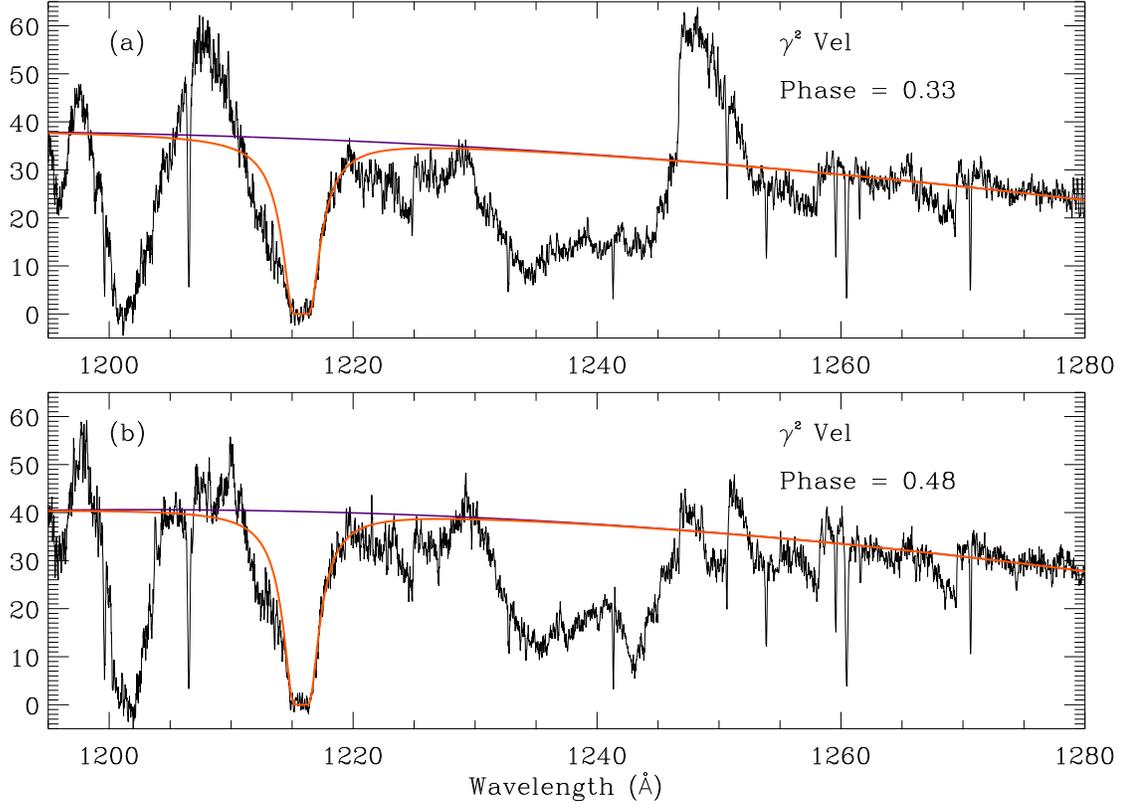}{12cm}{90}{68}{68}{270}{0}
\caption[]{Samples of the \iue\ observations of \gvel\ 
used to derive the interstellar \nhi: (a) SWP6175, 
spectroscopic binary phase = 0.33, and (b) SWP4719, phase = 0.48 
(phases were calculated using the period and $T_{0}$ from Schmutz et 
al.\markcite{sch97} 1997). These spectra have been smoothed with a 
5-pixel boxcar for display purposes only; the unsmoothed data were used 
to constrain \nhi\ as described in the text. The best-fitting 
\hi\ profiles (dotted lines) and continua (dashed lines) 
are overplotted on the data. Note the 
dramatic variability of the P-Cygni profiles. Note also the presence of 
complicated absorption structure which makes the blue wing of the 
\lya\ profile difficult to use for constraining \nhi. 
Both of these observations were obtained with the small \iue\ 
aperture.\label{gvHIsamples}}
\end{figure}

\placefigure{gvHIsamples} 

As noted in \S 1, \gvel\ is a complex stellar system with a strongly variable UV stellar spectrum. Figure \ref{gvHIsamples} shows two examples of the 
high-dispersion \iue\ spectra of \gvel\ that we have used to derive \nhi. 
These examples were selected to illustrate the variability of the 
P-Cygni emission lines in the vicinity of \lya, which are weakest at orbital 
phase $\sim$0.5. Figure \ref{gvHIsamples} also shows the complex and variable 
stellar absorption superposed on much of the blue wing of the interstellar \lya\ profile.  
This variability was a source of concern for this 
paper. Can we derive reliable \hi\ column densities despite the complex 
spectral changes occurring in this star?  Due to the stellar absorption, the 
blue wing of \lya\ was not useful for constraining \nhi. Fortunately, the 
red wing of \lya\ was clean and stable as a function of orbital phase. This can 
be seen by comparing panel (a) to panel (b) in Figure \ref{gvHIsamples}. In this paper, we used only the red wing for fitting the \lya\ profile, and the 
resultant fits generally look quite good (see Figures \ref{gvHIsamples} and 
\ref{gvHIblowup}).

We examined \nhi\ as a function of orbital phase of the spectroscopic binary to check for systematic errors in the derived interstellar \hi\ column density due to the complicated variability of the stellar spectrum.  There are considerable discrepancies in the orbital elements 
published by various groups (Stickland \& Lloyd 1990; Schmutz et 
al. 1997, and references therein). We adopt the orbital elements 
derived by Schmutz et al. (1997) who find a period of $78\fd53 
\pm0\fd01$ d with velocity semiamplitudes of $K_{\rm WR} = 122\pm2 \kms$ and 
$K_{\rm O} = 38.4\pm 2 \kms$.  In addition to the usual intrinsic variability of WR stars, it is believed that the \gvel\ spectrum may also change as a result of periodic absorption of the O star component by the WR wind (Stickland \& Lloyd 1990).

\begin{figure}
%\plotone{fig8_gvhiphase-lin.ps}
\plotone{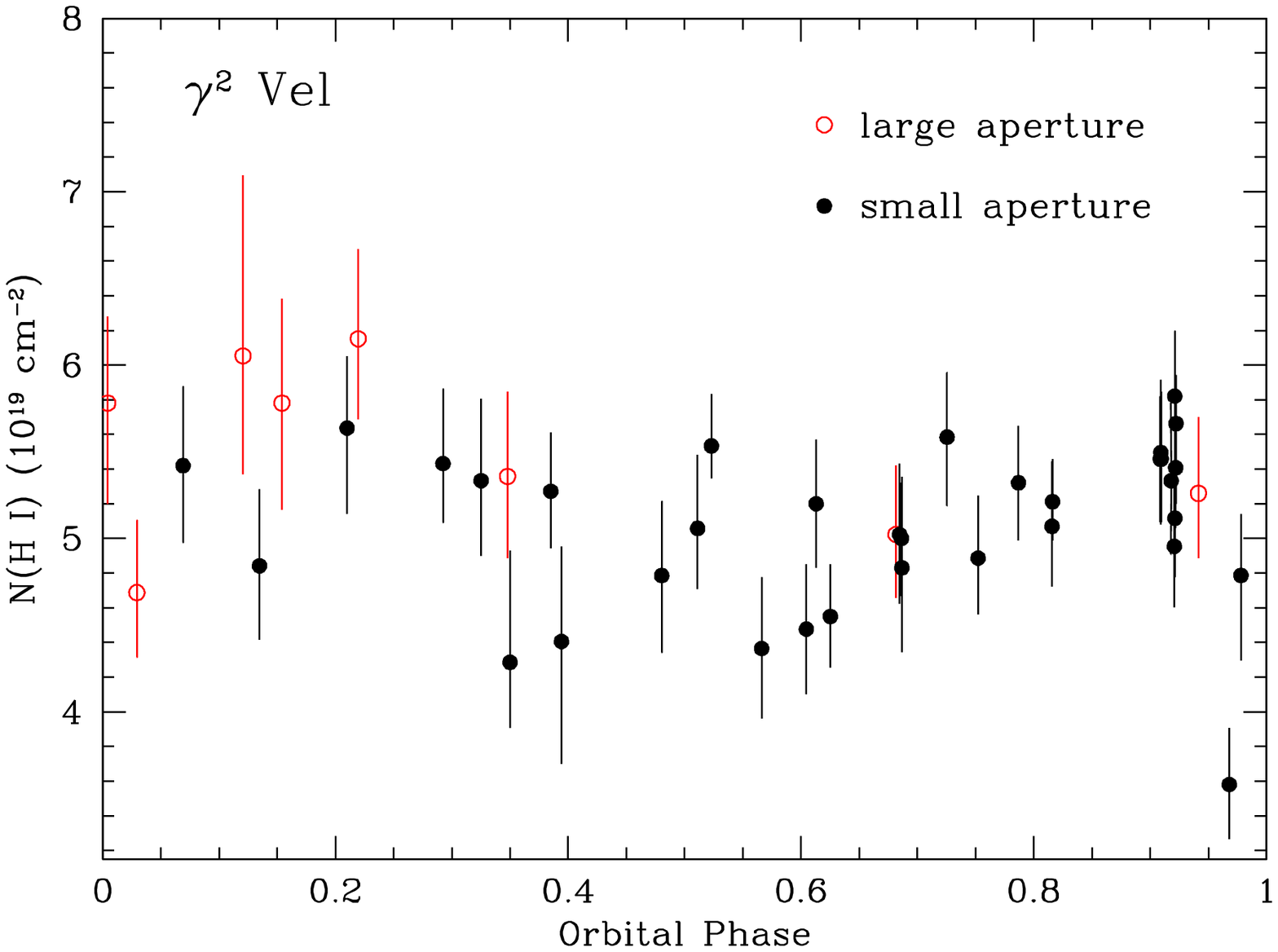}
\caption[]{Interstellar \hi\ column densities derived from the 42 
\iue\ observations of \gvel, plotted versus phase of 
the spectroscopic binary. Each column density is plotted with $\pm 
1\sigma$ error bars. Data obtained with the large aperture are 
indicated with open circles while small aperture results are shown with 
filled circles.\label{gvHIfig}}
\end{figure}

After screening the \iue\ high-dispersion observations of \gvel, we were left 
with 42 spectra, 34 
small aperture observations obtained early in the \iue\ mission and 8 
later observations obtained with the large aperture. In Figure 
\ref{gvHIfig} we plot the \hi\ column densities derived from all 
of the observations as a function of the phase of the spectroscopic 
binary, and Table \ref{g2nhi} summarizes the $<$\nhi$>$, 
$\sigma$, and $\epsilon$ derived from the large aperture data only, the 
small aperture data only, and all of the data combined. While one might 
argue that a trend is apparent in Figure \ref{gvHIfig} with a minimum 
in the derived \nhi\ values at orbital phase $\sim$ 0.5, this 
is a marginal result at best, and we do not believe that it should be 
taken too seriously. On the contrary, the generally good agreement of 
the \hi\ column densities at various phases despite the dramatic 
changes in the stellar spectrum (see Figure \ref{gvHIsamples}) is 
encouraging, and we conclude from Figure \ref{gvHIfig} that the \iue\ 
data provide a good determination of $<$\nhi$>$. Since the small and large 
aperture measurements are in good agreement, we adopted the results 
from the combined data set. Systematic errors probably cause the true 
uncertainty in $<$\nhi$>$ to exceed the formal error estimate provided in Table 
\ref{g2nhi}. Drawing from our experience with the more extensive data set for 
\zpup, we conservatively adopt 0.0256 dex as the $1\sigma$ uncertainty in \nhi\ 
toward \gvel\ as well: \nhi$=5.13\pm0.30\EE{19}\cmsq$.

\placefigure{gvHIfig}

\placetable{g2nhi}
\begin{deluxetable}{lccccc}
\tablewidth{0pc}
\tablecaption{Interstellar \nhi Toward \gvel\label{g2nhi}}
\tablehead{Data & No. & $<$\nhi$>$ & $\sigma$\tablenotemark{b} & 
$\epsilon$\tablenotemark{c} & $\chi^{2}_{\nu}$\tablenotemark{~d} \nl
Set\tablenotemark{a} & Spectra & ($\cmsq$) & ($\cmsq$) & 
($\cmsq$) & \ }
\startdata
Lg. Aperture & 8 & 5.35$\times 10^{19}$ & 0.51$\times 10^{19}$ & 
0.18$\times 10^{19}$ & 1.15 \nl
Sm. Aperture & 34 & 5.10$\times 10^{19}$ & 0.48$\times 10^{19}$ & 
0.08$\times 10^{19}$ & 1.71 \nl
All Data       & 42 & 5.13$\times 10^{19}$ & 0.49$\times 10^{19}$ & 
0.08$\times 10^{19}$ & 1.62
\enddata
\tablenotetext{a}{All \nhi\ measurements were derived from 
\iue\  high dispersion spectra. This table shows the mean \hi\ column density, 
$<$\nhi$>$, derived from the large \iue\ aperture data only, the 
small aperture data only, and all of the data combined.}
\tablenotetext{b}{$\sigma$ = rms dispersion (both $\sigma$ and 
$<$\nhi$>$ were weighted inversely by the variances of the individual 
\nhi\ measurements).}
\tablenotetext{c}{$\epsilon$ = error in the mean = $\sigma$/(No. 
measurements)$^{0.5}$.}
\tablenotetext{d}{Reduced $\chi ^{2}_{\nu} = \chi ^{2}$/(degrees of 
freedom) ,where $\chi ^{2}=\sum_i \left\{ \left[ N_i({\rm H~I})-<N({\rm 
H~I})>\right]/\sigma\left[ N({\rm H~I})\right]_i \right\}^2$.}
\end{deluxetable}
\clearpage

\section{DEUTERIUM AND NITROGEN ABUNDANCE RATIOS}\label{D/H}
\subsection{D/H}
Combining the results for \di\ and \hi\ column densities derived in the previous 
sections, we determined the atomic D/H abundance ratio for \gvel\ and \zpup.   
Toward \gvel\ we found that D/H $=2.18^{+0.36}_{-0.31} \EE{-5}$ (the errors are 
90\% confidence limits). This result is a large departure from the value of D/H 
usually attributed as ``typical'' in the ISM of the Galaxy ($\sim1.5\EE{-5}$).  A large value for the average D/H ratio implies 
that even larger values could exist in individual components if other components 
have lower D/H values closer to the LISM ratio. Although the previous D/H
measurement for \gvel\ ($=2.0^{+1.1}_{-0.7}\EE{-5}$, York \& Rogerson 1976) appears to be in close agreement with the IMAPS result, we consider this to be  coincidental in view of the magnitude of the uncertainties in the \oao\ measurement.

For \zpup\ we find \ndi\ $=1.30^{+0.19}_{-0.17} \EE{15} \cmsq$, 
a value slightly below the lower limit 
derived from \oao\ spectra by Vidal-Madjar et 
al. (1977). They found that a large range in \ndi\ ($1.5 - 20. \EE{15} \cmsq$) was possible for this sight line due to its complexity and a lack of adequate constraints.  Since the value of \nhi\ derived by Vidal-Madjar et al. (1977) for \zpup\ is consistent with ours, their large range in D/H must be a result of the \ndi\ uncertainty.  Much more is known now about the complexity of this
sightline than at the time of the \oao\ study 
(e.g. Welty, Morton \& Hobbs 1996).  More importantly, 
the IMAPS spectra have sufficient spectral resolution that the structure of the \zpup\ sightline in neutral hydrogen can be defined by the \nav\ profile for \ni\ and used to constrain the determination of \ndi. Thus with higher spectral 
resolution and better knowledge of the velocity structure along the sightline, a more accurate value of \ndi\ and D/H is derived.  We find D/H
$=1.42^{+0.25}_{-0.23} \EE{-5}$. 
The \ndi, \nhi, and D/H results for the three stars studied by IMAPS are 
summarized in Table \ref{abund3}, where all errors are presented as 90\% confidence limits. 

\placetable{abund3}
\begin{deluxetable}{cccc}
\tablewidth{0pc}
\tablecaption{Abundance Ratios\tablenotemark{a} \label{abund3}}
\tablehead{~ & \gvel\ & \zpup\ & \dori\ 
 \ }
\startdata
\ndi\ $(10^{15} \cmsq)$ & $1.12^{+0.15}_{-0.12}$ & $1.30^{+0.19}_{-0.17}$ & 
$1.16^{+0.29}_{-0.20}$ \nl
$N$(\ni) $(10^{15} \cmsq)$ & $4.10\pm0.34$ & $7.62\pm0.84$ & $6.19\pm0.51$ \nl
\nhi\ $(10^{19} \cmsq)$ & $5.13\pm0.50$ & $9.18\pm0.89$ & 
 $15.6\pm1.52$\tablenotemark{b} \nl
D/H $(10^{-5})$ & $2.18^{+0.36}_{-0.31}$ & $1.42^{+0.25}_{-0.23}$ & 
$0.74^{+0.20}_{-0.15}$\tablenotemark{b} \nl
N/H $(10^{-5})$ & $7.99\pm1.02$ & $8.30\pm1.22$ & $3.97\pm0.51$ \nl
D/N             & $0.27\pm0.04$ & $0.17\pm0.03$ & $0.19\pm0.05$ \nl
\enddata
\tablenotetext{a} {All errors presented in this table are 90\% confidence limits 
(1.65$\sigma$).}
\tablenotetext{b} {Errors in \nhi\ for \dori\ from \pap1 were revised to reflect the fact that potential systematic effects in \nhi\ 
identified in this paper for \zpup\ may dominate the statistical errors quoted in the earlier paper.  The $1\sigma$ error in \nhi\ was set to 0.0256 dex. The errors in D/H for \dori\ are essentially the same as found in \pap1\ since they are dominated by the errors in \ndi.}
\end{deluxetable}

\begin{figure}
\plotfiddle{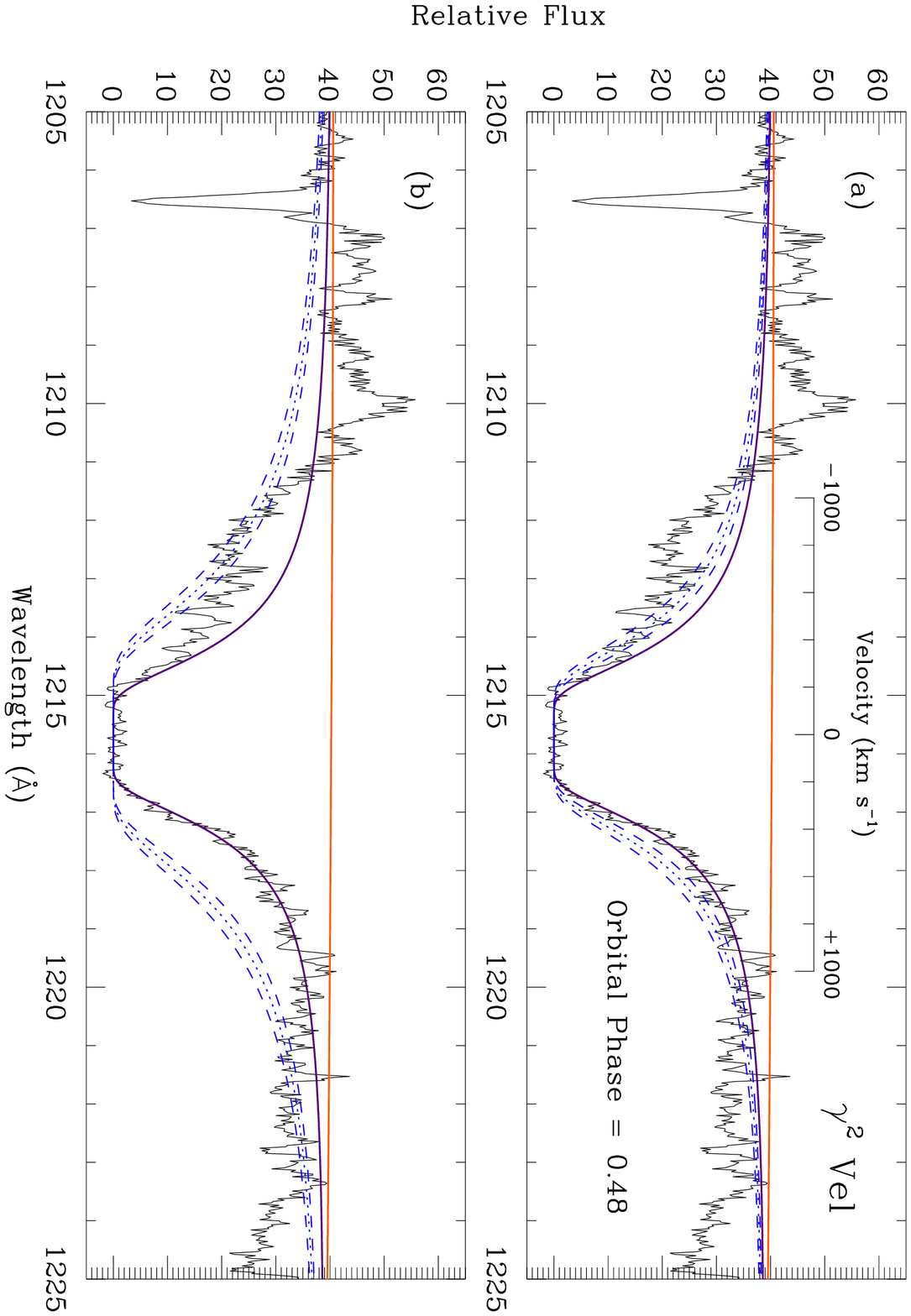}{12cm}{90}{68}{68}{270}{0}
%\plotfiddle{fig9_gvhi2.ps}{12cm}{90}{68}{68}{270}{0}
\caption[]{Expanded plots of the \hi\ \lya\ profile in \gvel\ shown in 
Figure \ref{gvHIsamples}b (SWP4719), again smoothed with a 5-pixel 
boxcar for display purposes only. The best-fitting profile and its 
corresponding continuum are indicated with solid lines. The broader 
damping profiles outside the best fit show the appearance of the 
profile when \nhi\ is forced to take on a value so that (a) 
D/H $= 1.5\EE{-5}$ as generally observed in the local ISM, given 
the values of \ndi\ in Table \ref{zpdtab} [the dotted line assumes the 
most probable \ndi\ while the dashed lines use the 90\% 
confidence limits on \ndi], and (b) D/H = $0.74\EE{-
5}$ as observed toward \dori\ in \pap1. Note that the region 
between 1207 and 1210 \AA\ is sometimes affected by P-Cygni emission (which 
happens to be weak in this observation, see Figure 
\ref{gvHIsamples}).\label{gvHIblowup}}
\end{figure}

\placefigure{gvHIblowup}

We now address a critical question, namely, is there sufficient uncertainty in 
the \nhi\ and \ndi\
measurements to reconcile the \gvel\ D/H abundance ratio with 
the general result observed in the local ISM or with the D/H derived in 
\pap1\ for the sight line to \dori? Figure \ref{gvHIblowup}a  
shows with dotted and dashed lines the expected appearance the 
\lya\ profile would have if our measurement of \ndi\ is 
correct but the \nhi\ were made high enough so that D/H $=1.5\EE{-5}$. In Figure \ref{gvHIblowup}, the dotted line represents the \hi\ profile that would
be needed for the most probable value for 
\ndi, while the dashed lines represent the 90\% confidence limits. For 
comparison, we also show in Figure \ref{gvHIblowup}a the best-fitting \hi\ 
profile (solid line) derived in the previous section for this particular 
observation. Figure \ref{gvHIblowup}b is an analogous plot with \nhi\ forced 
to take on a value so that D/H $=0.74\EE{-5}$ as observed toward 
\dori\ (\pap1). From Figure \ref{gvHIblowup} it is clear that forcing \nhi\ 
to make D/H 
$=1.5\EE{-5}$, assuming \ndi\ from \S\ref{gvelN(DI)}, produces a 
very poor (and unacceptable) fit to the \hi\ profile, and the fit 
becomes ridiculous if D/H $=0.74 \EE{-5}$.  We have assumed the 
best-fit continuum (solid line) to compute the \lya\
profiles shown in Figure \ref{gvHIblowup}. We have also explored 
whether or not the D/H ratios can be reconciled by adjusting the 
continuum placement, and we found that absurd continuum placements and shapes 
must 
be assumed to make D/H $=1.5\EE{-5}$ or (even worse) $0.74 \EE{-5}$. 

As demonstrated at the end of \S\ref{gvelN(DI)}, the changes in \ndi\ needed to make D/H $=1.5\EE{-5}$ or $=0.74 \EE{-5}$, assuming our 
most probable value of \nhi, gave a large mismatch with the data, which were clearly unacceptable. We conclude that spatial inhomogeneities in D/H in 
the ISM within $\sim$500 pc of the Sun have high statistical significance, 
especially when contrasting the D/H ratios in the directions of \gvel\ and
\dori, which were measured with the same instrument and techniques.

\subsection{Nitrogen Abundances}

The \ni\ column density \nni\ was computed by integrating the column density 
profiles shown in Figure \ref{NI_Navfig}. The \nni results for the three targets 
are listed in Table \ref{abund3}.  This table also contains the resulting N/H 
and D/N abundance ratios.  We examined the potential error in \nni\ by examining 
the errors in the portions of the various \ni\ profiles used to construct \nav.   
The  $1\sigma$ uncertainty in \nni\ for \gvel\ and \dori\ is 
conservatively estimated to be 5\% 
since the \ni\ profiles were all of high quality.  There may be 
some question about the structure in the core of \nav\ for \zpup. 
The uncertainty in \nni for \zpup\ 
was estimated by supposing the spike in \nav\ at $v \sim 19 \kms$ is due largely 
to noise fluctuations in the core of \ni\ \wl 952.5, the weakest \ni\ line we 
detect on this sightline.  If we truncate the  \nav\ profile at the level of the 
secondary maximum at $v\sim 24 \kms$ [\nav$=5.6\EE{14}\cmsq (\kms)^{-1}$], we 
find that the area of the spike is $3.6\EE{14} \cmsq$, or 4.7\% of the total.  
We then suppose that the total uncertainty  in \nni\ for \zpup\ is $\sqrt{2}$ 
times this, or 1$\sigma=5.1\EE{14}\cmsq$, to allow for the possibility that 
there may be other such fluctuations.  

\section{DISCUSSION}\label{discuss}

\begin{figure}
\plotone{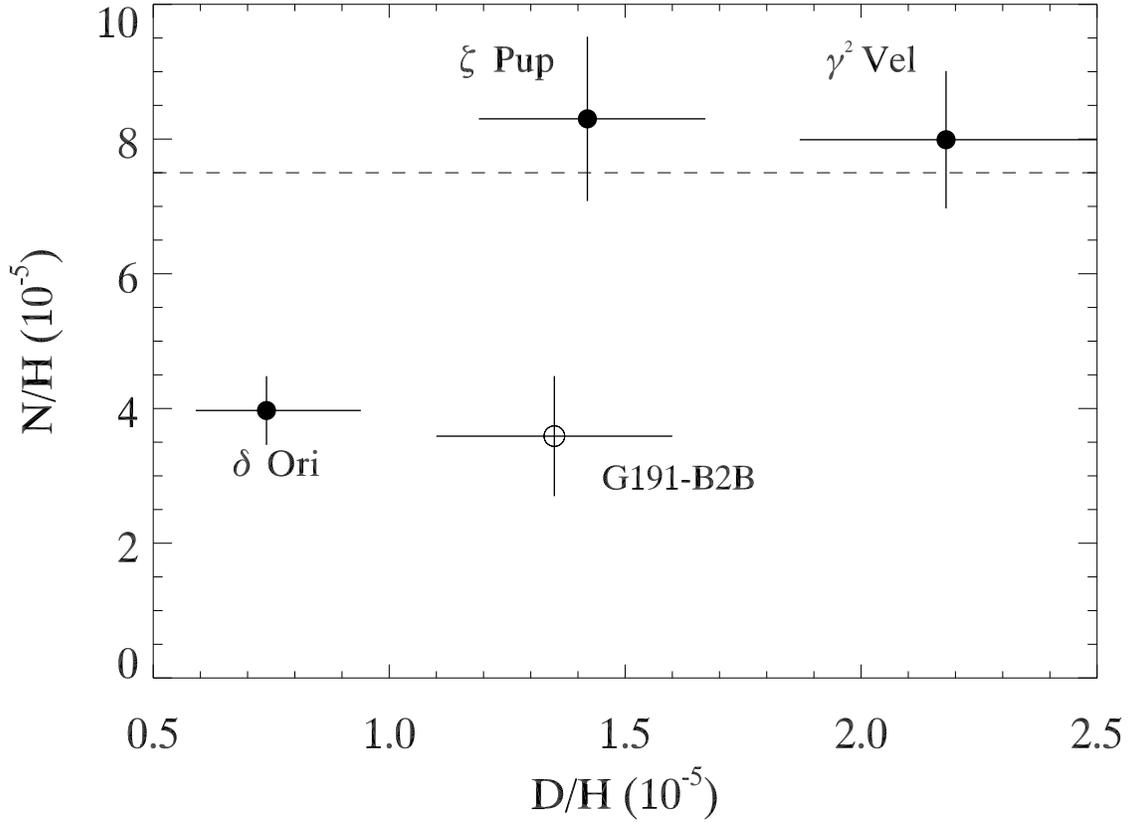}
\caption[]{$N/H$ abundance ratio as a function of $D/H$ for the three stars 
studied by IMAPS plus the white dwarf G191-B2B. The abundance ratios for G191-B2B represent the full range of values reported by Vidal-Madjar et al. (1998) and Sahu et al. (1999).  The average value of N/H in the ISM found by Meyer, Cardelli, \& Sofia (1997) is indicated by a dashed line.  
The errors bars are 90\% confidence limits (1.65$\sigma$). \label{dnfig}}
\end{figure}

\placefigure{dnfig}

There are two principal conclusions of the IMAPS D/H program. First, the atomic 
D/H ratio in the ISM, averaged over path lengths of 250 to 500 pc, exhibits 
significant spatial variability. Differences in the atomic D/H ratio on long pathlengths in the ISM have been suspected for 
many years (Vidal-Madjar et al. 1978; Vidal-Madjar \& Gry 1984), but not until now have data of sufficient quality been available to 
evaluate and reduce statistical and systematic errors to levels where these 
differences are unequivocal. Second, we find no support for the simple picture that variations in D/H
anticorrelate with those of N/H, i.e., one measure of how much the gas has been
processed through stellar interiors.  Figure \ref{dnfig} shows the relationship
between D/H and N/H for the three sightlines studied by IMAPS plus the white dwarf
G191$-$B2B (Vidal-Madjar et al. 1998; Sahu et al. 1999).  We point out that some
elements are systematically removed from the gas phase as they are incorporated
into interstellar dust (Savage \& Sembach 1996), but the abundance of N does not
seem to be appreciably altered by this effect (Meyer et al. 1997).

Beyond the effects from depletions onto dust, spatial variations in interstellar
gas abundances can arise as a natural consequence of galactic chemical evolution
and the changing influences of different stellar populations.  The lack of an
anticorrelation between N/H and D/H depicted in Fig. \ref{dnfig} indicates that
the variability of D/H is not just a consequence of different mixing ratios of
material with differing levels of stellar processing, as we might anticipate, for
instance, from the variable addition of metal-poor, infalling gas from the
Galactic halo (Meyer et al. 1994).  We may need to go further and draw a
distinction between contributions from stars that simply destroy deuterium and
those that both destroy deuterium and enrich the medium with additional nitrogen. That is, we could envision some stars cycling material only through their shallow
layers that are only hot enough to burn deuterium, while others eject material
from much deeper layers where the synthesis of heavier elements has taken place. 
This additional level of complexity could explain the behavior that we observed.

Global models of Galactic chemical evolution (Audouze \& Tinsley 1974; Tosi 1988a,b; Dearborn, Steigman \& Tosi 1996; Scully et al. 1996; Tosi et al. 1998) describe the destruction of D during stellar formation, evolution, and eventual mass loss. These models predict variations in D/H, N, and O abundances that are manifested as 
abundance gradients on a scale of $\ga$1 kpc. 
However, the predicted trends in galactic abundances may not accurately represent what is observable in the diffuse ISM.  Tenorio-Tagle (1996) showed 
that the chemical enrichment of the diffuse ISM by OB associations,  
including supernovae from massive stars, is a slow process.  Following a 
supernova explosion, chemically enriched ejecta remain clumpy and not well mixed with 
the diffuse ISM it encounters until it is incorporated into new star forming regions. This is a result of very long time scales for diffusion between different parcels of gas in the warm ($10^4$K) and cold ($10^2$K) phases of the ISM.

The diffusion time scale for enriched gas to thoroughly mix with the warm diffuse ISM can be very long ($t_D > 10^{10}$ yr - Tenorio-Tagle 1996). On the spatial scale  
sampled by the IMAPS observations, different sight lines may encounter regions with very different dynamical and chemical histories. 
Differential galactic rotation and random 
cloud motions are expected to stir the diffuse ISM and chemically enriched 
parcels of gas, but these parcels retain their distinct chemical properties until they 
are disrupted through photoevaporation, most likely by the formation of new massive stars. Tenorio-Tagle finds that diffusion is efficient only for the
hot phase of the ISM ($t_D<10^6$ yr), which accounts for a very small fraction of the total diffuse ISM. Only after the 
enriched gas and diffuse gas are highly
ionized would the chemically enriched gas from the earlier generation of stars quickly diffuse into the ambient ISM.   
Thus, the time scale for mixing interstellar gases
with different processing histories can be much longer than the chemical
evolution time scale. However, it is possible that interstellar turbulence  and its secondary phenomena may accelerate the mixing rate.
Since the distribution of star forming regions (OB associations) shows large 
inhomogeneities on scales $\la1$ kpc, their 
corresponding chemical enrichment of the ISM may be expected to be nonuniform
as well.   This is perhaps revealed indirectly by variations in \hii\
region abundances (Peimbert 1999) and solar-type stars at
similar galactocentric radii (Edvardsson, et al. 1993).

Several processes unrelated to stellar nucleosynthesis may, under the right circumstances,
alter the atomic D/H ratio of some parcels of interstellar gas (see Lemoine et al. 1999 for a review).  D may be incorporated 
into HD (Watson 1973), but the fraction of molecular gas on our sight lines is very low (see \pap1). Differential radiation pressure on D and H (Vidal-Madjar 
et al. 1978; Bruston et al. 1981) may lead to a separation of D in some clouds 
near strong radiation fields. Adsorption of D onto dust grains (Jura 1982) may 
deplete D from the gas phase.   Bauschlicher (1998) found that 
reactions of H and D with polycyclic aromatic hydrocarbon (PAH) cations might 
systematically provide some D enrichment in PAHs.  
A very different perspective has been offered by Mullan \& 
Linsky (1999), who  suggested that significant quantities of D may be formed in 
stellar flares from M dwarf stars and ejected into interstellar space.
Some or all of these 
processes could be at work in the diffuse ISM and alter the atomic D/H ratio on 
individual sight lines, independent of the degree of chemical enrichment from stellar evolution.  However, they have not yet been demonstrated to be quantitatively significant.
Observational and theoretical tests of the efficiency and 
applicability of these 
processes are needed to better understand the mechanisms 
affecting the D/H ratio in the diffuse ISM.

While the D/H ratios derived from IMAPS spectra are in the general range 
expected from galactic chemical evolution models, a factor of 
three variation in the mean D/H ratio on path lengths of several hundred pc is 
unexpected. The apparent lack of an anti-correlation of the D/H abundance ratio with the metallicity of 
the gas and its variability on smaller than expected scales suggests that other 
processes in the Galaxy may be masking more general chemical evolution 
trends. This may pose a problem for deriving a ``primordial'' D/H by 
extrapolating back from D/H measurements in the Milky Way to extragalactic 
absorbers at higher and higher redshifts, or to even to evaluate ``primordial'' D/H directly from high redshift observations, until we understand the reasons for these differences.  
The spatial variations found in this study underline the importance of 
high-quality D/H determinations.  D/H measurements in more distant regions
of the Galaxy are needed to determine whether 
the properties of the gas within 
500 pc of the Sun are representative of the Galactic disk.  
Observations with the  FUSE satellite should probe such more distant
environments and hopefully answer some of these questions.

The value of D/H for \gvel\ is larger than that usually considered typical for the Milky Way.  This robust result establishes a new lower limit to the primordial D/H ratio.   Within the framework of standard Big Bang Nucleosynthesis (Walker et al. 1991), the large value of D/H found toward \gvel\ is equivalent to a cosmic baryon density of $\Omega_B h^2 = 0.023\pm0.002$.  This error simply reflects the uncertainty in the D/H ratio toward \gvel\ reported in this paper.  We regard this value of $\Omega_B h^2$ as an upper limit since no correction has been applied for the destruction of deuterium in stars. This upper limit on $\Omega_B h^{2}$ is consistent with the preferred values of $\Omega_B h^{2}$ derived from recent analyses of the BOOMERANG and MAXIMA cosmic microwave background measurements (e.g., Lange et al. 2000; Tegmark \& Zaldarriaga 2000; Hu et al. 2000). However, any lowering of this upper limit on $\Omega_B h^{2}$ to correct for astration will lead to a marginal disagreement with simple inflation models (see Figure 4 in Tegmark \& Zaldarriaga 2000) and requires adjustments of other cosmological parameters. Alternatively, the $\Omega_B h^{2}$ upper limit from D/H may be taken as a prior assumption for the constraint of other cosmological parameters using the CMB data (e.g., see Hu et al. 2000).

We wish to thank the US and German space agencies, NASA and DARA, for their 
joint support of the \orf\ mission that made these observations possible. The 
successful execution of our observations was the product of efforts over many 
years by engineering teams at Princeton University Observatory, Ball Aerospace 
Systems Group, and Daimler-Benz Aerospace. Important contributions 
to the success of IMAPS also came from the efforts of D. A. Content and 
R. A. Keski-Kuha and other
members of the Optics Branch of the NASA 
Goddard Space Flight Center and from O. H. Siegmund and S. R. Jelinsky at the 
Berkeley Space Sciences Laboratory. We also thank Bruce Draine for insightful 
discussions on atomic interactions with dust grains and PAHs.  This work was 
supported in part by NASA grant NAG5-616 to Princeton University. 
The \iue\ data 
were obtained from the National Space Science Data Center at NASA-Goddard.


\begin{references}
\reference{ag89} Anders, E., \& Grevesse, N. 1989, Geochim. Cosmochim. Acta, 53, 197
\reference{at74} Audouze, J., \& Tinsley, B. 1974, ApJ, 192, 487
\reference{bau98} Bauschlicher, C. W. 1998, ApJ, 509, L125
\reference{br92} Bevington, P. R., \& Robinson, D. K. 1992, Data Reduction and 
Error Analysis for the Physical Sciences, 2nd ed. (New York: McGraw Hill)
\reference{bb84} Bianchi, L., \& Bohlin, R. C. 1984, A\&A, 134, 31
\reference{bluhm99} Bluhm, H., Marggraf, O., de Boer, K. S., Richter, P., \& Heber, U. 1999, A\&A, 352, 287
\reference{BS85} Boesgaard, A. M., \& Steigman, G. 1985, ARAA, 23, 319
\reference{bsd78} Bohlin, R. C., Savage, B. D., \& Drake, J. F. 1978, 
ApJ, 224, 132
\reference{brus81} Bruston, P., Audouze, J., Vidal-Madjar, A., \& Laurent, C. 
1981, ApJ, 243, 161
\reference{dst96} Dearborn, D. S. P., Steigman, G., \& Tosi, M. 1996, ApJ, 465, 887
\reference{ds94} Diplas, A., \& Savage, B. D. 1994, ApJS, 93, 211
\reference{edv93} Edvardsson, B., Anderson, J., Gustafsson, B., Lambert, D. L., Nissen, P. E., Tomkin, J. 1993, A\&A, 275, 101
\reference{els74} Epstein, R. L., Latimer, J., \& Schramm, D. N. 1976, 
Nature, 263, 198
\reference{f81} Ferlet, R. 1981, A\&A, 98, L1
\reference{fs94} Fitzpatrick, E. L., \& Spitzer, L. 1994, ApJ, 427, 232
\reference{gold92} Goldbach, C., L\"udtke, T., Martin, M., \& Nollez, G. 1992, A\&A, 266, 605
\reference{g98} G\"olz, M., et al. 1998, in Proc. IAU Colloq. No. 166, The Local Bubble and Beyond, eds. D. Breitschwerdt, M. J. Freyberg, J. Tr\"umper (Berlin: Springer), 75
\reference{bsc 82} Hoffleit, D. \& Jaschek, C. 1982, The Bright Star Catalogue,  4th ed., (New Haven: Yale U. Obs.)
\reference{hpm95} Howarth, I. D., Prinja, R. K., \& Massa, D. 1995, 
ApJ, 452, L65
\reference{hs2000} Howk, J. C., \& Sembach, K. R. 2000, AJ, in press, 
astro-ph/9912388
\reference{hu00} Hu., W., Fukugita, M., Zaldarriaga, M., \& Tegmark, M. 2000, astro-ph/0006436
\reference{hur98} Hurwitz, M., et al. 1998, ApJ, 500, L1
\reference{j71} Jenkins, E. B. 1971, ApJ, 169, 25
\reference{jp97} Jenkins, E. B., \& Peimbert, A. 1997, ApJ, 477, 265
\reference{j88} Jenkins, E. B., Joseph, C. L., Long, D., Zucchino, P. M., Carruthers, G. R., Bottema, M., \& Delamere, W. A. 1988, in  UV Technology II, ed. R. E. Huffman (Bellingham: International Society for Optical Engineering), 213
\reference{j96b} Jenkins, E. B., Reale, M. A., Zucchino, P. M., \& Sofia, U. J.  1996, Ap\&SS, 239, 315
\reference{j00} Jenkins, E. B. et al., 2000, ApJ Letters, in press
\reference{pap1} Jenkins, E.B., Tripp, T. M., Wo\'zniak, P. R., Sofia, U. 
J., \& Sonneborn, G. 1999a, ApJ, 520, 182 (Paper I)
\reference{j99b} Jenkins, E. B., et al. 1999b, BAAS, 31, 942
\reference{jura82} Jura, M., 1982, in Advances in Ultraviolet Astronomy (NASA CP-2238) ed. Y. Kondo (Greenbelt: NASA), 54
\reference{lmb76} Lampton, M., Margon, B., \& Bowyer, S. 1976, ApJ, 208, 177
\reference{lang00} Lange, A. E., et al. 2000, astro-ph/0005004
\reference{Lem99} Lemoine, M., et al. 1999, New Astr., 4, 231
\reference{l93} Linsky, J. L. et al. 1993, ApJ, 402, 694
\reference{l95} Linsky, J. L., Diplas, A., Wood, B. E., Brown, A., Ayres, T. R., \& Savage, B. D. 1995, ApJ, 451, 335
\reference{m95} Massa, D. et al. 1995, ApJ, 452, L53
\reference{m98} Massa, D., Van Steenberg, M. E., Oliversen, N., \& 
Lawton, P. 1998, in Ultraviolet Astrophysics Beyond the \iue\ Final 
Archive (ESA SP-413), ed. W. Wamsteker \& R. Gonzalez Riestra (Noordwijk: ESA), 723
\reference{meyer97} Meyer, D. M., Cardelli, J. A., \& Sofia, U. J. 1997, ApJ, 490, L103
\reference{meyer94} Meyer, D. M., Jura, M., Hawkins, I., \& Cardelli, J. A. 1994, ApJ, 437, L59
\reference{moos00} Moos, H. W., et al. 2000, ApJ Letters, in press
\reference{mort78} Morton, D. C. 1978, ApJ, 222, 863
\reference{mortb79} Morton, D. C., \& Bhavsar, S. P. 1979, ApJ, 228, 147
\reference{mortd76} Morton, D. C. \& Dinerstein, H. L. 1976, ApJ, 204, 1
\reference{mortu77} Morton, D. C, \& Underhill, A. B. 1977, ApJS, 33, 83
\reference{mull99} Mullan, D. J., \& Linsky, J. L. 1999, ApJ, 511, 502
\reference{peim99} Peimbert, M. 1999, in Chemical Evolution from Zero to High Redshift, eds. J. R. Walsh \& M. R. Rosa (Berlin: Springer), 30
\reference{press92} Press, W. H., et al. 1992, Numerical Recipes in Fortran, 2nd ed. (Cambridge: Cambridge Univ. Press)
\reference{p92} Prinja, R. K. et al. 1992, ApJ, 390, 266
\reference{pul94} Puldrach, A. W. A., Kudritzki, R. P., Puls, J., Butler, K., \& Hunsinger, J. 1994, A\&A, 283, 525
\reference{rh96} Reid, A. H. N., \& Howarth, I. D. 1996, A\&A 311, 616
\reference{oao3} Rogerson, J. B., Spitzer, L., Drake, J. F., Dressler, K., Jenkins, E. B., Morton, D. C., \& York, D. G. 1973, ApJ, L97
\reference{ry73} Rogerson, J. B., \& York, D. G. 1973, ApJ, 186, 
L95
\reference{sahu92} Sahu, M. S. 1992, Ph.D. Thesis, Univ. Groningen
\reference{sahu99} Sahu, M. S. et al. 1999, ApJ, 523, L159
\reference{ss91} Savage, B. D. \& Sembach, K. R. 1996, ARAA, 34, 279
\reference{hipp97} Schaerer, D., Schmutz, W., \& Grenon, M. 1997, ApJ, 
484, L153
\reference{sch97} Schmutz, W. et al. 1997, A\&A, 328, 219
\reference{xx} Scully, S. T., Cass\'e, M., Olive, K. A., Schramm, D. N.,
Truran, J., \& Vangoini-Flam, E. 1996, ApJ, 462, 960
\reference{ss85} Shull, J. M., \& Van Steenberg, M. E. 1985, ApJ, 294, 599
\reference{uj98} Sofia, U. J., \& Jenkins, E. B. 1998, ApJ, 499, 951
\reference{sl90} Strickland, D. J., \& Lloyd, C. 1990, Observatory, 
110, 1
\reference{teg00} Tegmark, M., \& Zaldarriaga, M. 2000, astro-ph/0004393 v3
\reference{tt96} Tenorio-Tagle, G. 1996, AJ, 111, 1641
\reference{tosi88a} Tosi, M. 1988a, A\&A, 197, 33
\reference{tosi88b} Tosi, M. 1988b, A\&A, 197, 47
\reference{tosi98} Tosi, M., Steigman, G., Matteucci, F., \& Chiappini, C.  1998, ApJ, 498, 226
\reference{vdh81} van der Hucht, K. A., Conti, P. S., Lundstr\"{o}m, I., 
\& Stenholm, B. 1981, Space Sci. Rev. 28, 307
\reference{vm00} Vidal-Madjar, A. 2000, in The Light Elements and Their 
Evolution, eds. L. da Silva, M. Spite, \& J. R. de Medieros, A.S.P. Conf. Ser., 
in press
\reference{vmg84} Vidal-Madjar, A., \& Gry, C. 1984, A\&A, 138, 285
\reference{vm77} Vidal-Madjar, A., Laurent, C., Bonnet, R. M., \& 
York, D. G. 1977, ApJ, 211, 91
\reference{vm78} Vidal-Madjar, A., Laurent, C., Bruston, P., \& Audouze, J. 
1978, ApJ, 223, 589
\reference{vm98} Vidal-Madjar, A., et al. 1998, A\&A, 338, 694
\reference{walk91} Walker, T. P., Steigman, G., Schramm, D. N., 
Olive, K. A., \& Kang, H.-S. 1991, ApJ, 376, 51
\reference{wat73} Watson, W. D., 1973, ApJ, 182, L73
\reference{wmh96} Welty. D. E., Morton, D. C., \& Hobbs, L. M. 1996, ApJS, 106, 
533
\reference{yr76} York, D. G., \& Rogerson, J. B. 1976, ApJ, 203, 378
\reference{y83} York, D. G., Spitzer, L., Bohlin, R. C., Hill, J., Jenkins, E. B., Savage, B. D., \& Snow, T. P. 1983, ApJ, 266, L55
\end{references}
\end{document}